# A De-biased Direct Question Approach to Measuring Consumers' Willingness to Pay

Reto Hofstetter
Professor of Marketing
Faculty of Economics and Management, University of Lucerne
Frohburgstrasse 3, P.O. Box 4466, 6002 Lucerne, Switzerland
Phone: (41) 41 229 58 80
Email: reto.hofstetter@unilu.ch

Klaus M. Miller
Assistant Professor of Quantitative Marketing
Faculty of Economics and Business Administration, Goethe University Frankfurt
Theodor-W.-Adorno-Platz 4, 60629 Frankfurt am Main
Phone: (49) 69 798 34601
Email: klaus.miller@wiwi.uni-frankfurt.de

Harley Krohmer
Professor of Marketing
University of Bern, Engehaldenstrasse 4, 3012 Bern, Switzerland
Phone: (41) 31 631 8031
Fax: (41) 31 631 8032
Email: krohmer@imu.unibe.ch

Z. John Zhang*
Tsai Wan-Tsai Professor, Professor of Marketing
Director, Penn Wharton China Center
The Wharton School
University of Pennsylvania, 3730 Walnut St, Philadelphia, PA 19104, USA
Phone: (01) 215 898 1989
Fax: (01) 215 898 2534
Email: zjzhang@wharton.upenn.edu

* We thank Bernd Skiera, Shunyao Yan, Rahul Thondan, and seminar participants at Goethe University Frankfurt, University of Geneva, Vienna University of Economics and Business, Bocconi University, and the University of Berne, as well as participants at JPIM Research Forum, PSI, EMAC, and AMA conferences and two anonymous IJRM reviewers, the Area Editor, and Werner Reinartz for helpful comments and feedback. This paper is based on the first two authors' dissertation, who are listed in alphabetical order and contributed equally to this paper. All remaining errors are our own.

# A De-biased Direct Question Approach to Measuring Consumers' Willingness to Pay

Knowledge of consumers' willingness to pay (WTP) is a prerequisite to profitable price-setting. To gauge consumers' WTP, practitioners often rely on a direct single question approach in which consumers are asked to explicitly state their WTP for a product. Despite its popularity among practitioners, this approach has been found to suffer from hypothetical bias. In this paper, we propose a rigorous method that improves the accuracy of the direct single question approach. Specifically, we systematically assess the hypothetical biases associated with the direct single question approach and explore ways to de-bias it. Our results show that by using the de-biasing procedures we propose, we can generate a de-biased direct single question approach that is accurate enough to be useful for managerial decision-making. We validate this approach with two studies in this paper.

The key to the optimal pricing decision for new and existing products and services is an accurate understanding of consumers' willingness to pay[1] (WTP; Anderson, Jain, and Chintagunta 1993). Consumers' WTP is also important for implementing various pricing tactics, such as nonlinear pricing (Jedidi and Zhang 2002), one-to-one pricing (Shaffer and Zhang 2000), and targeted promotions (Shaffer and Zhang 1995). Not surprisingly, various approaches have been developed to determine consumers' WTP (Miller et al. 2011). The main distinctions among these approaches are whether they measure WTP directly or indirectly, whether they use a single question or multiple questions, and whether they determine consumers' actual or hypothetical WTP (see Table 1 for an overview of the various methods to measure WTP and Table 2 for a structured comparison of the advantages and disadvantages of each approach).

< INSERT TABLE 1 ABOUT HERE >

< INSERT TABLE 2 ABOUT HERE >

In practice, market researchers often ask consumers to state their WTP for a product directly (Anderson, Jain, and Chintagunta 1992; Steiner and Hendus 2012; Hofstetter et al. 2013). This can be done using either a single, open-ended (OE) question format (Arrow et al. 1993; Mitchell and Carson 2013) or multiple, open-ended questions such as in the Van Westendrop Method (VWM; Van Westendorp 1976). In a management survey we conducted of 82 pricing managers, we found that the direct approach is the most popular approach used to determine demand (used by 28%, see Web Appendix A). The direct approach is also widely used by market research firms. Steiner and Hendus (2012) find that 76% of the surveyed firms use a direct approach. The enduring popularity of the direct approach is due to its obvious advantages. Conceptually simple and easy to implement with regard to data collection and analysis, the direct approach unfailingly generates timely information at a low cost (Jedidi and Jagpal 2009). Advances in digital technologies today make the direct approach seem to shine even more brightly because it facilitates massive online collections of consumer WTP data about a large number of products within a very short time[2]. The popularity of the direct approach is also helped by its inclusion in commercial applications, e.g., the price sensitivity meter (PSM and PSM plus) from GfK and BASES Price Advisor from Nielsen.

---

[1] In this paper, we take the standard economic view of consumer willingness to pay and define it as the maximum price at or below which a consumer will definitely buy one unit of the product. This corresponds to the concept of the floor reservation price as proposed by Wang, Venkatesh, and Chatterjee (2007). However, we do not adopt their idea of conceptualizing WTP as a range. Instead, we consider WTP as a point measure, staying in line with earlier literature in economics on measuring consumer WTP (e.g., Wertenbroch and Skiera 2002; Miller et al. 2011) and account for the individual variation of consumer WTP by constructing appropriate confidence intervals for our WTP measures at the aggregate level (see results section below). Further, we focus on the WTP for a product as a whole, assuming known availability and awareness of the product. Our study does not address the WTP for features of a product.

An alternative way to measure WTP is to use the indirect approach, such as using a single dichotomous-choice (DC) question format (Mitchell and Carson 2013), or multiple sequential questions such as in a choice-based conjoint analysis (CBC; Louviere and Woodworth 1983). The indirect approach has been shown to capture more realistic choice and purchase scenarios (Leigh, MacKay, and Summers 1984) and can, as in the case of conjoint analysis, provide additional information on the WTP for individual product attributes (e.g., Hanson and Martin 1990; Green and Krieger 1996; Jedidi and Zhang 2002). The downside of these indirect approaches is that except perhaps for the DC question format, they all require more effort in data collection and more expertise in analysis. The effort and expertise required can be quite costly and discouraging for many practitioners, so much so that many shun those approaches.

One thing that both the direct and indirect approaches have in common is the fact that they all elicit consumers' hypothetical WTP (HWTP), because they do not typically require the respondents to actually buy the product (Wertenbroch and Skiera 2002; Ding, Grewal, and Liechty 2005; Miller et al. 2011). Hypothetical WTP, which corresponds to consumers' stated preferences, can deviate from their actual WTP (AWTP; Hoffmann et al. 1993), which reflects their revealed preferences. This deviation, known in the economics literature as hypothetical bias, is induced by the hypothetical nature of a task (Harrison and Rutström 2008). One way to remove this bias is to use incentive-aligned direct methods to measure consumers' actual WTP. The BDM mechanism using a single question as proposed by Becker, DeGroot, and Marschak (BDM; 1964) is one such method, and the ICERANGE approach using multiple questions as suggested by Wang, Venkatesh, and Chatterjee (2007) is another. A consumer's actual WTP can also be measured indirectly using the incentive-aligned counterpart of the single dichotomous choice (DC) question format (Brynjolfsson, Collis, and Eggers 2019) or multiple-question, incentive-aligned, choice-based conjoint analysis (ICBC; Ding, Grewal, and Liechty 2005; Ding 2007; Dong, Ding, and Huber 2010)[3].

---

[2] See also Brynjolfsson, Collis, and Eggers (2019) who use an indirect, single question approach, the dichotomous-choice (DC) question format, to collect massive online choice experimental data to measure changes in well-being. In a similar vein, open-ended (OE) questions can easily be administered to a large number of respondents via the internet.

For researchers and practitioners, an incentive-aligned approach should be the method of choice. However, an incentive-aligned approach may not always be feasible in the case where product prototypes are not available or privacy concerns and legal restrictions in some countries prevent their usage, or the incentives are too costly to provide or to simulate. Indeed, Europe's General Data Protection Regulation (GDPR) has rendered incentive-aligned methods that require the collection of respondents' personal data (e.g., email addresses, phone numbers, etc.) much more difficult and costly to implement. Even if incentive-aligned approaches are feasible, their application may be very costly, for example, for high-value items such as a car or a house. Further, incentive-aligned approaches put an excessive burden on the respondents in terms of understanding the research method as well as the required time to respond (Wertenbroch and Skiera 2002; Ding, Grewal, and Liechty 2005; Ding 2007; Miller et al. 2011).[4] For all these reasons, scholars and practitioners are looking hard for new ways to remove the hypothetical bias from non-incentive aligned approaches. Interestingly, in that pursuit, most scholars focus on the more complex, indirect approach--conjoint analysis--and have had a number of successes with that method (see e.g., Jedidi and Jagpal 2009). These successes include formulas related to data calibration (Brazell et al. 2006), mental simulation (Hoeffler 2003), and mixing hypothetical choice data with a small amount of incentive-aligned choice experimental data (Laghaie and Otter 2019).

[3] Other incentive-aligned approaches include a sequential incentive-compatible conjoint procedure for eliciting consumer WTP for attribute upgrades proposed by Park, Ding, and Rao (2008) or an incentive-compatible dynamic auction for selling multiple complementary goods as suggested by Sun and Yang (2014).

In contrast, academic researchers pay scant attention to simpler, direct approaches such as the OE question format, even though practitioners routinely use such an approach. In this paper, we will develop some rigorous de-biasing procedures for OE, the simplest direct approach in use. Our de-biasing strategy is to calibrate data from two single question formats, OE and DC, as illustrated in Figure 1[5], in a theory-informed way. Specifically, we systematically investigate the nature of hypothetical biases associated with the data from two single question formats in a marketing context based on past research and leverage the inherent bias structure associated with each question format to improve the data from OE. We show that by using the de-biasing procedures we propose here, the use of data from OE can help practitioners make more accurate managerial decisions without sacrificing the simplicity and timeliness they value (see Web Appendix I for a managerial guide on how to apply the de-biasing approach in market research practice). We demonstrate the value of our proposed de-biasing procedures through two online studies. In each study, our procedures perform remarkably well.

[4] We note that in practice, market researchers can also obtain consumers' actual WTP from real market transactions such as field experiments, scanner data, online purchases, or simulated test market data. Actual WTP data from past transactions is incentive compatible and shows a high convergent validity due to actual purchase observations under realistic market conditions. However, because consumers' true WTP remains unknown, the interpretation of these data is difficult (Wertenbroch and Skiera 2002). Also, there is not always sufficient natural variation in prices or only within a very limited range for the focal firm's product and its competitive products to estimate the true WTP (Jedidi and Jagpal 2009). Field experiments, most notably online these days, can provide some remedy, but similar to the various sources of transaction data mentioned above, they are not feasible for new products or are simply too expensive to implement.

[5] Some studies in the experimental economics literature refer to a third single question format, the payment-card (PC) question format (see Mitchell and Carson 2013 for an overview). With the PC question format, respondents are asked to select one of several proposed prices as their maximum WTP for a specific product. If their maximum WTP differs from their proposed WTP, respondents can still give an individual value by using an open-ended field. We did not include this format in our study because it is in fact a hybrid question format that combines elements of price generation (OE) and price selection tasks (DC).

< INSERT FIGURE 1 ABOUT HERE >

Previous studies have documented the existence of the hypothetical bias for various question formats (e.g., Bishop, Welsh, and Heberlein 1992; Balistreri et al. 2001; Harrison and Rutström 2008) and the extent of this bias (Lusk and Schroeder 2004; Miller et al. 2011; Schmidt and Bijmolt 2019). A few researchers have looked specifically into which single question format yields the least biased (i.e., most valid) results by comparing two subsets of single question approaches, the DC and OE formats (Bishop, Welsh, and Heberlein 1992; Brown et al. 1996; Loomis et al. 1997; Balistreri et al. 2001; Murphy et al. 2005). A number of ex-ante and ex-post calibration techniques have been proposed for de-biasing single question-based pricing surveys (Carson 2000; Loomis 2011). Ex-ante techniques attempt to improve hypothetical methods at the data collection stage through priming survey subjects, whereas ex-post approaches try to calibrate data after measuring WTP. These techniques all rest on the assumption that, although responses to hypothetical questions may be biased, these responses provide useful information if extracted properly (Murphy and Stevens 2004).

Ex-ante de-biasing techniques include letting survey subjects understand the consequences of their answers (Carson, Groves, and List 2006; Cummings, Harrison, and Osborne 1995; Landry and List, 2007), urging their honesty and realism before the survey (Loomis, González-Cabán, and Robin 1996; Jacquemet et al. 2013; Stevens, Tabatabaei, and Lass 2013), or reminding them of possible biases (Cummings and Taylor 1999; List 2001; Poe et al. 2002; Adland and Caplan 2003; Brown, Ajzen, and Hrubes 2003; Lusk 2003). These prior studies have shown mixed results in eliminating the hypothetical bias, which is plausible given that consumers may not always adjust their actual behaviors based on verbal reminders alone (e.g., Farrell and Rabin 1996).

Ex-post de-biasing approaches include methods such as uncertainty adjustment (e.g., Champ et al. 1997; Champ and Bishop 2001; Poe et al. 2002), the Bayesian Truth Serum (e.g., Prelec 2004; Weaver and Prelec 2013), and response calibration (e.g. Arrow et al. 1993; Fox et al. 1998; List and Shogren 2002; Hofler and List 2004; Murphy and Stevens 2004; Murphy et al. 2005). Ex-post de-biasing approaches can yield good approximations of actual payments in some applications, but not in others. It remains unclear, however, how well these ex-post de-biasing approaches would work in our specific context of measuring WTP for a consumer products, since they were developed in other contexts such as measuring WTP for non-market, public goods.

Our study differs from those cited above in three significant ways: First, we conduct our investigation in the context of pricing a regular consumer good that marketing practitioners typically deal with, not in the context of a public good. Second, we appeal to the past research to identify the theoretical bias structure associated with the data from each question format. Then, we collect single-source data to directly compare consumers' actual willingness to pay (AWTP) from BDM to the WTP information solicited through the single question format (i.e., OE and DC). This enables us to verify the theoretical bias structure of each single question format. Third, we base our de-biasing procedures on theory. We do so by analytically leveraging the bias structures inherent in each question format and identifying the ideal de-biasing procedure for the data elicited through OE. Based on that ideal procedure, we then propose three levels of precision in de-biasing for practitioners and validate our proposed procedures by applying them to the single-source data collected. In short, our research provides some practical de-biasing procedures for the direct question approach that are conceptually sound and complete with promising external validity.

It is important to note that our research here is not meant to promote the usage of the direct question approach or to pass judgment on the adequacy or inadequacy of alternative approaches such as conjoint analysis. Rather, it is to improve the accuracy of the direct question approach, if practitioners and researchers choose to use them for one reason or another.

## *MODEL DEVELOPMENT*

In this section, we formalize a bias model for a price generation task (i.e., the OE question format) as well as for a price selection task (i.e., the DC question format). We will subsequently use these models to derive a robust de-biasing formula for estimating true WTP based on data collected through OE and DC questions. See Table 3 for a summary of our model notation and definition of the respective variables.

Let $p_i$ be the actual WTP for an individual consumer $i$ where $p_i$ is a random draw from $f(p_i)$ with $p_i \sim N(\bar{p}, \sigma_{p_i}^2)$ and $p_i \geq 0$. Further, let $\tilde{p}_i$ be the stated WTP for an individual consumer $i$ when stating her maximum WTP under an OE question format and $\hat{p}_i$ under a DC question format. Similarly, we define respective means as $\overline{p} = \frac{1}{n}\sum_{i=1}^{n} p_i$, $\tilde{p} = \frac{1}{n}\sum_{i=1}^{n} \tilde{p}_i$, and $\hat{p} = \frac{1}{n}\sum_{i=1}^{n} \hat{p}_i$.

For each respondent $i$ in the two single question formats, we observe either $\tilde{p}_i$ or $\hat{p}_i$. If the respondent is unbiased, we should observe $p_i = \tilde{p}_i$ and $p_i = \hat{p}_i$, respectively. However, stated

WTP often deviates from actual WTP, as consumers often need to construct their WTP in response to the specific elicitation context rather than retrieving a previously formed value stored in their memory (Bettman, Luce, and Payne 1998; O'Donnell and Evers 2019). Past literature has shown multiple reasons for both price generation tasks and price selection tasks to generate biases specific to question formats. In the following, we will draw on this literature to derive question format-specific bias models for price generation (OE) and price selection (DC) tasks, which we will use subsequently to de-bias the data from the OE format.

< INSERT TABLE 3 ABOUT HERE >

***Biases in Price Generation Formats (OE Question Format).*** Price generation tasks, such as the OE question format, do not provide the respondent with any cues regarding reference prices or price ranges to the respondent and hence allow respondents an unlimited degree of freedom to state their WTP (Chernev 2003). There is little research on how consumers respond to price generation tasks (Chernev 2006), however, the little we do know indicates that consumers have to go through a three-stage price generation process. First, they need to evoke the range of possible values. Next, they use these values as benchmarks to determine the product's potential utility to themselves and articulate this utility on a monetary scale, thus forming their maximum WTP (Chernev 2003). Finally, when consumers state their true WTP at the moment of the survey, they need to discount their hypothetical future WTP (Loomis et al. 1997). As the effort in this process is generally not sufficiently rewarded (Kemp and Maxwell 1993; Arrow et al. 1993), respondents may not properly discount their WTP and hence report inflated values leading to an individual, consumer-level bias (List 2001; Ding, Grewal, and Liechty 2005; Miller et al. 2011). This inflated individual-specific bias is also confirmed by a study from Loomis et al. (1997). In that study, respondents typically inflate the amount they would pay in a hypothetical purchasing situation as compared to an incentive-aligned purchasing situation. This lack of incentive-compatibility leading to an over-stated WTP for price generation tasks is also noted by Hoehn and Randall (1987) and Carson, Groves, and Machina (2000). Therefore, we can write stated WTP $\tilde{p}_i$ from OE as

$$\tilde{p}_i = (\alpha + \varepsilon_i) + p_i. \tag{1}$$

In other words, a consumer's stated WTP is an inflated value of her true WTP and the inflation factor $\alpha + \varepsilon_i$ is individual-specific. Here, $\varepsilon_i \sim N(0, \sigma_{\varepsilon_i}^2)$ captures individual-specific variations and $\alpha$ is a product category-specific inflator that is typically positive (but is not required to be) as noted in Dickie, Fisher, and Gerking (1987), Shogren (1990), Carson et al. (1996), as well as List and Shogren (1998). In summary, Equation (1) describes the well-documented bias that an individual tends to inflate her true WTP with an inflator that varies by the individual, but will have a category-specific positive bias.

***Biases in Price Selection Formats (DC Question Format).*** Price selection tasks, such as the dichotomous-choice (DC) question format, lighten the burden of setting a monetary value for a product by presenting the consumer with a single price for the product under study. The consumer's decision in this case is reduced to whether to select the product at the given price.

It is well known that the price selection task has its own issues. According to Kahneman, Slovic, and Tversky (1982), people under uncertainties are prone, when making estimates, to start from an initial value, which they will then adjust to yield the final answer. Such an anchoring strategy will bias toward the initial value because the adjustments are typically insufficient (Slovic and Lichtenstein 1971). In price selection tasks, anchoring occurs when respondents seize upon the price cue as the value of the product (Mitchell and Carson 2013). Studies have shown that such anchoring is indeed common. The respondent confronted with a dollar figure in a situation where she is uncertain about a product's value, the respondent may regard the proposed amount as conveying approximation of the product's true value. She will then anchor her WTP on the proposed amount (Green, Jacowitz, Kahneman, and McFadden 1998; Mitchell and Carson 2013). That is, respondents have a propensity to inflate their true WTP when faced with a high price cue. Conversely, when confronted with a low price cue, respondents tend to underreport their true WTP.

This tendency of anchoring to inflate or deflate actual WTP is also rooted in the perception that price and quality are often correlated, to the extent that a price cue may be taken as an indicator of product quality (Gabor and Granger 1966; Shapiro 1968; Monroe 1973). According to the literature on the price-quality relationship, consumers who make price-quality inferences tend to prefer higher-priced products when price is the only information available, as they apparently perceive the more expensive product to be of higher quality (Monroe 1973; John, Scott, and Bettman 1986). Consumers in this situation may be more likely to accept prices greater than their actual WTP. This may explain so-called *yea-saying* (i.e., consumers accept a given price as their WTP although their actual WTP is lower), which has been observed in experimental economics (e.g., Brown et al. 1996; Mitchell and Carson 2013).

In contrast, if a relatively low price is presented to the consumer, she may perceive the product to be of low quality. In this case, consumers will be less likely to accept prices lower than their actual WTP. This phenomenon is known as *nay-saying*, that is, consumers reject a given price, although their actual WTP is greater than that amount (Carson 2000). Both yea- and nay-saying is more likely to occur the farther away the price cues are from a respondent’s actual WTP. These biases are likely smaller when the price cues are closer to her actual WTP (Green, Jacowitz, Kahneman, and McFadden 1998; Carson 2000; Carson, Groves, and Machina 2000; Mitchell and Carson 2013; Bishop et. al. 2017). The previous discussions suggest that we can specify a consumer’s stated WTP elicited through the dichotomous-choice format (DC) as:

$$\hat{p}_i = \theta_i(p_i^c - p_i) + p_i \quad (2)$$

where $\theta_i$ with a zero mean is the individual's specific bias under a dichotomous-choice (DC) format and $p_i^c$ is the price cue presented to consumer $i$. Equation (2) captures the fact that under DC, a respondent's bias is anchored on the price cue presented to her; a higher (lower) price cue than her actual WTP will lead her to state an inflated (deflated) WTP[6].

***De-biasing Approach.*** As stated in the introduction, both OE and DC are easy to implement in practice. But once both data series are collected, the question becomes, how can we use these data to make managerial decisions? In the previous discussion, we showed that neither data series is ideal because of its inherent biases. However, we show in this section that by leveraging the bias structures of these two data series, we can actually de-bias the data series elicited through the OE question format so that it becomes possible to make all relevant managerial decisions unhampered by data biases.

To see how we can de-bias the OE series, we note that from Equation (1), we have $\frac{1}{n}\sum_{i=1}^{n}\tilde{p}_i = \alpha + \frac{1}{n}\sum_{i=1}^{n}\varepsilon_i + \frac{1}{n}\sum_{i=1}^{n}p_i$. Given the zero mean for $\varepsilon$ and definitions in Table 3, this implies

$$\alpha = \tilde{p} - \bar{p}. \tag{3}$$

Then from Equation (2), once again using the definitions in Table 3, we have $\hat{p} = \frac{1}{n}\sum_{i=1}^{n}\theta_i\, p_i^c - \frac{1}{n}\sum_{i=1}^{n}\theta_i p_i + \bar{p}$. As price cues are randomly assigned to respondents, $p_i^c$ and $\theta_i$ are independent variables[7]. This implies $\frac{1}{n}\sum_{i=1}^{n}\theta_i\, p_i^c = 0$, and $\hat{p} = \bar{p} - \frac{1}{n}\sum_{i=1}^{n}\theta_i p_i$. Also, note $\frac{1}{n}\sum\theta_i\, p_i = cov\,(\theta, p)$ (see Web Appendix C), where $cov\,(\theta, p)$ is the covariance. We must have

$$\bar{p} = \hat{p} + cov(\theta, p). \tag{4}$$

---

[6] At the zero mean $\theta_i$ can still have a distribution where there are more incidences for positive values than for negative values, such that a respondent is more likely to be biased upward (downward) at a high (low) price cue (see Web Appendix B).

[7] Random assignment is a sufficient condition, but not a necessary one. The necessary and sufficient condition is $p_i^c$ is not assigned conditional specifically on $\theta_i$.

From (3) and (4), we arrive at

$$\alpha = \tilde{p} - \hat{p} - cov(\theta, p). \tag{5}$$

Finally, from (1) and (5), we can derive the bias correcting function below

$$p_i = \tilde{p}_i - \tilde{p} + \hat{p} + cov(\theta, p) - \varepsilon_i. \tag{6}$$

Equation (6) suggests that a researcher can derive a respondent's actual WTP from stated WTP elicited from the OE question format by adding three non-individual, specific adjustment factors: $\tilde{p}$, $\hat{p}$, and $cov(\theta, p)$ plus $\varepsilon_i$, a white noise. The first two factors are known from the two biased data series collected through OE and DC and the third factor can be simulated easily as it is a constant number.

This enabled us to conduct our de-biasing procedure for the OE question format in four steps. First, we collected the two biased data series. Second, we de-biased the OE series by using $\tilde{p}$ and $\hat{p}$ only in Equation (6). In other words, we set $cov(\theta, p) = 0$ and ignore $\varepsilon_i$ in Equation (6) as the first order approximation. We call this our BASIC de-biasing procedure. Third, we introduced $\varepsilon_i$ while still keeping $cov(\theta, p) = 0$. We refer to this as our EPSILON de-biasing procedure. Finally, we introduced both $cov(\theta, p) \neq 0$ and $\varepsilon_i$ . We call this our FULL de-biasing procedure. By following these four steps, we could tease out how effective each part of our de-biasing procedure for the OE data series is in helping us to make better managerial decisions. Finally, we check the results of these de-biasing procedures by measuring them against the gold standard of the consumer's actual WTP measurement, the output of the BDM mechanism (Becker, DeGroot, and Marschak 1964; Wertenbroch and Skiera 2002; Miller et al. 2011).

It is important to note that these proposed de-biasing procedures do not presume any knowledge of true WTP; we have collected the true WTP data through the BDM mechanism here merely to validate our procedures.

## *METHOD AND DATA COLLECTION*

We conducted two large-scale empirical studies to test our de-biasing procedures. These studies gauge WTP for two new products, a gym bag and a sweatshirt, among students at a major Swiss university. Both products were specifically produced for the purpose of this study and new to the target market at the time of the study. We report the gym bag study (Study 1) here in detail and give a summary of the sweatshirt study (Study 2).

***Participants***. The data for the gym bag study was collected through an online experiment. To recruit participants, we sent out 12,448 invitation emails to the entire student body (undergraduate, graduate, and Ph.D. candidates) of a large Swiss university. We motivated participation by offering all survey participants a chance to win an Apple iPhone 7 Plus in a raffle[8]. The participants were further informed that their chance to win the raffle was independent of their experimental responses[9]. A total of 826 participants chose to take part in the bag study, which represents a response rate of 6.64%. We pre-specified that data collection would end after seven days (i.e., the decision to stop collecting data was independent of the experimental results; we did not analyze the data until after data collection had been completed). Within the seven-day period, we collected as much data as we could.

***Stimulus.*** The stimulus we used in our study was a gym bag imprinted with a logo of the university that was not available in the market at the time of the study (see Web Appendix D for a depiction of the stimulus). The gym bag was designed and fabricated exclusively for the purpose of the study. We expected the university gym bag to be both interesting and affordable for most of the students, who actually represent the target segment of the product. Since apparel and accessories are also often sold online, the online channel represents the known and accustomed distribution channel for these categories[10]. Further, because the students had no reference for the exact market price for this distinctive university gym bag, their hypothetical WTP or actual WTP statements would not be capped[11].

---

[8] We used the smartphone as a single incentive to motivate participation in our survey in order to recruit an adequate number of subjects. However, the smartphone was not connected to our stimulus and the incentive-aligned condition under BDM, where proper incentives are offered to the participants so that they are motivated to reveal their true preferences (see Wertenbroch and Skiera 2002 for details).

[9] It is possible that some consumers may have taken part in the survey just to win the smartphone and may not have been interested in purchasing the gym bag.

As our stimulus was new to the market, no repeat purchases were observed and participants had never stated their WTP for the product before. Further, the university gym bag was not displayed in a competitive setting. As a result, students were unable to select the stimulus from a group of competing products as they would in a real online store. Finally, because we were conducting an online experiment, shipping the product had to be easy and cheap, which was the case with the gym bag.

***Experimental Design.*** We developed three different independent experimental groups and used a between-subjects design. Each participant was randomly assigned to one treatment group.

In the open-ended (OE) question format group, each participant directly stated her hypothetical WTP for the university gym bag.

In the dichotomous-choice (DC) question format group, we used a total of 21 price levels, ranging from Swiss Franc (CHF[12]) 1.25 up to CHF 26.25 incremented in steps of CHF 1.25. We chose the market price of the most expensive gym bag similar to our products as our upper limit of CHF 26.25. Each respondent received a price level that was chosen randomly from the 21 available levels. The random distribution of the price levels was even, meaning that all the price levels had an equal 1 in 21 chance of appearing in the respondent's DC question[13].

---

[10] See e.g., the Stanford Bookstore online: https://www.bkstr.com/stanfordstore/home.
[11] We acknowledge, however, that some participants may have a reference price from similar products in mind.
[12] At the time when this paper was written 1 CHF represents approximately 1 USD.

In the BDM group, which is our control group for validating our de-biasing procedures, we determined our benchmarking actual WTP data by using an incentive-aligned mechanism, the BDM mechanism proposed by Becker, DeGroot, and Marschak (1964). We chose the BDM mechanism as WTP from BDM has been found to not significantly differ from a consumer's WTP based on real purchase data (Miller et al. 2011). We implemented the BDM mechanism in a way similar to what Wertenbroch and Skiera (2002) did. In our particular application of the BDM mechanism, we told participants that they would have a chance to purchase the university gym bag without having to invest more money than they would be willing to pay for the product. We also informed them that the price for the university gym bag was not yet set, and that it would be determined randomly from a predefined uniform distribution unknown to the participants. Participants were further told that they were obligated to buy the university gym bag at the randomly determined price if the price was less than or equal to their stated WTP. However, if the randomly determined price was higher, a respondent would not have a chance to buy the product. This mechanism ensures that participants have no incentive to state a price that is higher or lower than their true WTP.

---

[13] In Web Appendix H, we test the robustness of our de-biasing approaches to the chosen DC price range in a simulation study. We find that the range of prices chosen in DC matters. If the DC range chosen is larger than the range of true WTP, reducing the DC range will not have any impact on our de-biasing procedures. However, if the DC range is contained by the true WTP, our de-biasing procedure is still robust, if we do not reduce the DC range by more than 35% of the true price range. Beyond that, our de-biasing procedures do not work as well and some work better than others. The de-biasing procedures do not work as well because an unrealistic range of prices in DC will simply distort the estimation of mean WTP. The simulation shows that the DC range needs to be carefully chosen to reflect the true minimum and maximum WTP in the market and that it is better to err on the wide side than on the narrow side of the DC range. In addition, we show that a parametric approach can help reduce biases if the DC range chosen is too narrow. We find the parametric approach to be highly robust to misspecifications of the DC range.

To carry out the buying obligation, we recorded the name and address of each participant in the BDM group. After the completion of the study, we determined each individual participant's buying obligation by drawing from a discrete uniform distribution which corresponded to the price-levels used under the DC question format. The distribution thus ranged from CHF 1.25 to CHF 26.25 incremented in steps of CHF 1.25 and included a total of 21 price levels. We determined per participant whether the randomly drawn price was smaller or equal to her stated WTP. Thus, none of the participants had to purchase the gym bag at a price that was larger than their stated WTP. Out of all participants in the BDM group, 19.85% were obliged to buy the product at an average price of 8.03 (SD = 5.66, min = 1.25, max = 26.25). Only 12 participants (21.81%) of those who were obliged to buy paid a price higher than the average BDM WTP of 13.04. After the completion of the study, all participants who were obligated to buy were sent the gym bag with an invoice, via the post. The invoice was due within 14 days and payable with cash or credit card (this payment process was officially approved by the appropriate university authorities). Out of all 277 respondents in the BDM group, 55 (19.86%) were required to purchase the bag. Only one respondent refused to comply with her purchase obligation and returned the product[14].

We obtained 270 responses in the OE group, 279 responses in the DC group, and 277 responses in the BDM group. Our realized sample size exceeds current expectations in experimental studies of larger than 50 respondents per cell (Simmons 2014).

[14] Valid actual WTP estimation requires that the respondents understand the BDM procedure and the underlying buying process. In our sample, respondents understood the BDM mechanism quite well. As Wertenbroch and Skiera (2002) and Miller et al. (2011) did, we asked the subjects if it was clear to them why it was in their best interest to state exactly the price they were willing to pay. Using a seven-point Likert scale from one (not at all) to seven (very much so), the participants responded with an average of 5.954. We used a similar method to determine the understanding of the buying process and found an average of 6.222. Finally, we asked respondents if stating their WTP for the product was a task which was easy to understand and complete and participants replied with an average of 5.851. (see Web Appendix E for the exact wording).

The three experimental groups did not differ significantly in terms of socio-demographics or socioeconomic status. We performed a multivariate analysis of variance (Pillai-Spur: F = .922, p = .562) for age (F = .465, p = .707), sex (F = 2.077, p = .102), education (F = .139, p = .937), occupation (F = .593, p = .620), income (F = 2.028, p = .109), budget for clothing and accessories (F = .1886, p = .904), and purchase interest (F = .604, p = .613).

***Experimental Procedure***. We divided our online experiment into three parts. The first part described the product (i.e., the university gym bag) in the OE, DC, and BDM groups. The second part consisted of the WTP task in the different experimental treatment groups. In the third part of the online experiment, we conducted a brief survey on socio-demographics and -economics, and we made sure the participants understood the WTP elicitation method to which they were exposed (see Web Appendix E for further details).

***WTP Estimation Procedure.*** Figure 2 plots the observed demand in each treatment group. For the OE and BDM groups, we obtained each respondent's hypothetical WTP and actual WTP directly from the survey data and plotted demand $q(p)$ as the probability that a respondent's WTP is equal to or greater than a certain price $p$ using the demand function of the form $q(p) = Pr(WTP \geq p)$. For the DC group, we plotted the choice share for each price level. We determined the face validity of WTP measures by correlating elicited WTP with the respondent's purchase interest. We measured purchase interest itself by using a seven-point Likert scale ranging from one (low interest in the product) to seven (high interest in the product). Face validity is high for all methods because correlations are positive and significant (OE: r = .390, p < .001, BDM: r = .216, p = .001). We did not test the DC question format because hypothetical WTP data was not available on an individual level.

## ***RESULTS***

***Study 1: University Gym Bag.*** First, we eyeball the empirical demand functions that can be generated from the three experimental groups (Figure 2). We observe that OE demand (circles) is systematically higher compared to BDM (dots) at most price levels. For DC (diamonds), the function appears to be underestimating demand at low prices and overestimating demand at high prices. Both observations are consistent with the biases identified in the literature, which we discussed in the model development section.

< INSERT FIGURE 2 ABOUT HERE >

Next, we compare statistically both OE and DC data series to the BDM. As summarized in Table 4, we find the OE mean to be statistically different from BDM as indicated by a t-test [$t(543.92)_{\text{OE vs. BDM}} = 4.167$, $p < .001$] and non-overlapping 95%-confidence intervals. It is also different from BDM in terms of the distribution of the data series, as indicated by a KS-test and LR-test ($D_{\text{KS-Test, OE vs. BDM}} = .181$, $p < .001$; $D_{\text{LR-Test, OE vs. BDM}} = 41.615$, $p < .001$). However, the DC mean does not statistically differ from the BDM as indicated by overlapping 95%-confidence intervals, as we expected, but does differ in distribution ($D_{\text{LR-Test, DC vs. BDM}} = 18.383$, $p < .001$). Thus, our statistical analysis supports our model specifications for the biases identified in the literature for OE as well as DC data series.

We now show, in two steps, that by using our de-biasing procedure on the OE data series, we can, firstly, significantly improve its statistical fit with BDM and, secondly, demonstrate that the de-biased data series can help managers make better managerial decisions. Table 4 summarizes the results of our first step. In the Table 4, $M_{\text{BASIC}}$ is the mean of the data series we generated by applying the BASIC de-biasing procedure, i.e., subtracting $\tilde{p} - \hat{p}$ from $\tilde{p}_i$ while setting $cov(\theta, p) = 0$ and $\varepsilon_i = 0$ (see Equation 6). This is the most straightforward case of de-biasing. We also generated $M_{\text{EPSILON}}$, the mean of the data series where we subtract $\tilde{p} - \hat{p}$ from $\tilde{p}_i$ while setting $cov(\theta, p) = 0$ and adding $\varepsilon_i$. In this simulation, we randomly drew $\varepsilon_i$ from a normal distribution with zero mean and the same standard deviation as the OE data series $SD(\tilde{p}_i)$. Finally, $M_{FULL}$ is the mean of the data series we have generated by fully simulating Equation (6). Here, $cov(\theta, p)$ is simulated in the wide range of $[-3.08, 7.08]$, and in that range, the best theoretical ($cov(\theta, p) = 2.08$) and empirical ($cov(\theta, p) = 2.33$) outcomes for $M_{FULL}$ are reported in Table 4 (Later in the paper, on page 24 and following, we will elaborate on how to select this range of $cov(\theta, p)$ and determine its theoretical best value.)

From Table 4, we can see that our BASIC de-biasing procedure has generated a data series that is closer to the BDM data series. Unlike the original OE data series, the 95%-confidence interval of the mean from the de-biased data series now overlaps with that of the BDM, although the t-test [$t(543.94)_{BASIC\ vs.\ BDM} = 2.76$, $p < .01$] and the KS-test ($D_{KS\text{-}Test,\ BASIC\ vs.\ BDM} = .26$, $p < .01$) remain negative. When individual-specific variations are accounted for in generating $M_{EPSILON}$, we generate a data series that is even closer to the BDM. Table 4 shows that this data series differs from the BDM only in distribution as indicated by the negative KS-test ($D_{KS\text{-}Test,\ EPSILON\ vs.\ BDM} = .22$, $p < .01$), but not the mean [$t(513.5)_{EPSILON\ vs.\ BDM} = 1.20$, $p > .05$]. Finally, when both covariance and individual-specific variations are incorporated in the FULL approach, we similarly find that this data series differs from the BDM only in distribution as indicated by the negative KS-test when using our theoretically best value for $cov(\theta, p) = 2.08$ ([$t(505.31)_{FULL(cov=2.08)\ vs.\ BDM} = .97$, $p > .05$], $D_{KS\text{-}Test,\ FULL(cov=2.08)\ vs.\ BDM} = .14$, $p < .05$, as well as overlapping 95%-confidence intervals) and for our empirically best value for $cov(\theta, p) = 2.33$ [$t(502.03)_{FULL(cov=2.33)\ vs.\ BDM} = 1.03$, $p > .05$], $D_{KS\text{-}Test,\ FULL(cov=2.33)\ vs.\ BDM} = .16$, $p < .01$, as well as overlapping 95%-confidence intervals). Clearly, our de-biasing procedures have generated a data series that is statistically indistinguishable from the BDM as far as these two tests are concerned.

Of course, whether a de-biased data series is a winner will depend on how well it can help a firm make its managerial decisions. Here, we can use the optimal price, optimal quantity, and optimal profits generated from the BDM data as the benchmarks and then compare them to the same variables generated from the original OE and DC data as well as from the three de-biased data series discussed above (BASIC, EPSILON, and FULL). For instance, the BDM data series gives us the incentive-compatible relationships between price and sales quantity. For the manufacturer of the university gym bag, the profit function is $\pi = (p - c) \times q(p) \times ms$, where p is the price, c are the marginal costs, q(p) is quantity scaled from [0,1] given by $q(p) = {e^{\alpha+\beta\times p}}/{1+e^{\alpha+\beta\times p}}$ and ms is the market size. By incorporating the marginal cost for the university gym bag, which is c = CHF 5.00[15], and ms = 1,000 as the size of the target market, we can easily derive the optimal price, optimal quantity, and optimal profits[16]. Note that for the purpose of optimization, a nonzero fixed cost will not alter the outcome. Similarly, we can use the same cost and market size information for all other data series to generate the equivalent numbers, which we can then compare with those generated with the BDM data series. To make statistical comparisons, we use two measurements: overlapping confidence intervals for each variable comparison and the variable difference test.

To generate the confidence interval for an optimal variable from a data series, such as price, we bootstrap a data series 1,000 times and each time we generate an optimal price so that optimal prices from bootstrapping the data series generates a distribution for the optimal price (Efron and Tibshirani 1993). This way, we can compare the 95%-confidence interval from different data series to see if they overlap. If they do not, we conclude that the optimal prices from the two data series are different.

---

[15] According to the manufacturer, variable costs did not depend on the number of units produced. However, variable costs may only be constant over a certain range around the actual quantity ordered by the manufacturer.
[16] Since the university gym bag was not available for purchase elsewhere at the time of the study and due to the university branding did not have any direct competitors, we assumed a monopoly for the purpose of price optimization. In order to consider indirect competitors (e.g., other gym bag brands), researchers would need to estimate demand not only for one product, but also for all competitive products and allow for competitive equilibrium analysis.

To do the variable difference test, we once again bootstrap the two respective data series 1,000 times and each time we calculate the difference in an optimal variable from two data series (Efron and Tibshirani 1993). Thus, we generate a distribution of the differences of the two optimal variables. We then test if the mean difference in the optimal variable between the two bootstrapped data series is zero. If not, we conclude that the two optimal variables are statistically different.

In Table 5, we summarize the results of this analysis (for a visualization of the different demand curves resulting from the de-biasing approaches see Web Appendix Figure F.1). Relative to the outcomes using the BDM data, a firm would significantly inflate its estimates of the optimal quantity when using the original OE data, while the optimal price is higher but passes both tests. The consequence is that the firm's profit estimate based on OE data would be 38.22% higher than the estimate from the BDM data. This inflation can have dire consequences in the firm's decision making. The estimates from the DC data also show considerable biases in a much higher optimal price and a much lower optimal quantity, so that both variables fail the two tests. However, because of the compensating nature in the inflated optimal price and deflated optimal quantity, the estimate of the optimal profit would not be significantly different from the estimate of the BDM data by both tests. Of course, even if the biases cancel each other out to a degree and generate a good estimate of profitability, the use of the DC data can still have severe consequences for a firm in production and marketing decisions.

From Table 5, we can see that the two partial de-biasing procedures (BASIC and EPSILON) have both improved the estimates of the optimal price and optimal quantity, and both estimates pass the two tests. Relative to the original OE data, both data series have significantly improved the estimate of the optimal profits in absolute difference. Most importantly, with the FULL de-biasing procedure, our estimates of the optimal price, quantity, and profits all pass the two tests and our estimate of the optimal profit is less than 2% off the optimal profit drawn from BDM data. Our FULL de-biasing procedure has indeed delivered a rather sizeable improvement all around.

< INSERT TABLE 4 ABOUT HERE >

< INSERT TABLE 5 ABOUT HERE >

***Study 2: University Sweatshirt***. The previous gym bag study demonstrated the promise of our de-biasing procedure for the direct single question approach. In this second study, we applied the same procedure to a different data set we collected to investigate the robustness of our de-biasing approach. Here, we report a summary of the economic analysis of Study 2 and refer the reader to the detailed information on this study in the Web Appendix G.

As with the gym bag study, the OE data series on a university sweatshirt generate statistically different optimal quantity and the DC data series different optimal price estimates. Here too, the traditional interpretations of both data series lead to wildly overestimated profits, 48.83% and 23.03% respectively. The first BASIC de-biasing procedure subtracting only $\tilde{p} - \hat{p}_{.}$ improves significantly on the original OE data, but does not beat the DC data series. The second, EPSILON, where individual-specific variation is accounted for, improves on DC significantly and the profit estimate is different from that of BDM by only 6.16%. Our FULL de-biasing procedure performs, once again, remarkably well. The estimates of the optimal price and quantity are not statistically different from BDM and the estimate of optimal profits is within 5.26% of the BDM.

***Selecting the Right*** $\boldsymbol{cov(\theta, p)}$. From both studies, we have seen that our FULL de-biasing procedure statistically improved the OE data series in a robust way. More importantly, the de-biased data series using the FULL procedure always generated much improved estimates for the optimal price, quantity, and profits, judging by the gold standard of BDM estimates. These improvements can help a firm make better pricing decisions, better production planning, and better market entry decisions.

One remaining issue from both studies, which has great importance for researchers using our de-biasing procedure, is how to select a $cov(\theta, p)$. For our studies, we have simulated a wide range, $cov(\theta, p)\epsilon[-3.08, 7.08]$ in the first study and $cov(\theta, p)\epsilon[-8.20, 1.80]$ in the second study. As shown in Figure 3, our simulation results indicate that the value of $cov(\theta, p)$ that yields the profit estimate closest to the BDM is respectively 2.33 and -5.64. These numbers are, of course, not surprising to us, as we know from Equation (4) that the theoretical best value for $cov(\theta, p)$ for the purpose of our de-biasing procedure is given by $\bar{p} - \hat{p}$. In the first study, we have $\bar{p} - \hat{p}$ = 2.08 and in the second study, we have $\bar{p} - \hat{p}$ = -3.20. As Table 5 and Table F.2 show, the difference between the profit estimates at the empirical best $cov(\theta, p)$ values versus the theoretical best values is diminishingly small. Therefore, we suggest that analysts can use a small BDM sample to pin down $\bar{p}$ and then determine $cov(\theta, p)$ using DC data.

It is important to note that one needs a far smaller sample, only a handful of data points, to come up with an estimate of $\bar{p}$ (i.e., from BDM) than to estimate the optimal price, quantity, and profits. However, once a rough estimate of $\bar{p}$ is obtained, one can proceed to conduct sensitivity analysis around the $cov(\theta, p)$ implied by $\bar{p}$. The range of the $cov(\theta, p)$ used for the sensitivity analysis can be very small as the differences between the empirical best $cov(\theta, p)$ values and the theoretical best values are diminishingly small as noted earlier. In the case where even a small

BDM sample is too cumbersome to gather, our two studies confirm that a partial de-biasing procedure with $\tilde{p} - \hat{p}$ subtracted from the OE data series and accounting for individual-specific variations can still significantly improve our estimates relative to using the original OE data series.

< INSERT FIGURE 3 ABOUT HERE >

## *CONCLUSION*

In this paper, we take a close look at two single question approaches to gauge consumers' willingness to pay (WTP), the open-ended (OE) question and the dichotomous-choice (DC) question, which are frequently used by practitioners because the data are easy to collect. These approaches are conceptually simple, easy to implement, and unfailingly generate timely information, even in real-time, at a low cost. However, these advantages are negated if these approaches elicit a hypothetical WTP that deviates significantly from the consumers' actual WTP -- as is often the case. Marketing scholars have thus far failed to find ways to reduce hypothetical bias for single question approaches.

In this paper, we make the first attempt to de-bias the direct single question approach in a rigorous way. Specifically, we systematically investigate the nature of hypothetical biases associated with two basic single question formats, the open-ended (OE), and the dichotomous-choice (DC), and look for ways to leverage data from both question formats to reduce the bias in the OE question format. We do this by specifying an individual-level bias model based on theory and show that for price generation tasks such as the OE question format, a respondent tends to inflate her stated WTP, while for price selection tasks such as the DC question format, a respondent tends to show biases anchored on presented price cues. Further, we analytically derive a de-biasing procedure for the OE question format and show empirically the promise of this procedure.

Our results show that although the single question approach suffers from various kinds of statistical biases, there is no reason to discard the approach altogether. Indeed, when the market researcher uses the single question approach in combination with two question formats (i.e., the OE and DC question), she can overcome the specific biases generally associated with these formats by using our proposed de-biasing procedure. Our analysis shows that this procedure can de-bias the OE data enough to arrive at statistically and managerially valid forecasts of consumers' WTP without resorting to any BDM mechanism. When a small sample of BDM is obtainable, which is frequently the case in marketing applications, our proposed FULL procedure performs astonishingly well. Thus, our proposed de-biasing procedures preserve the advantages of simplicity and low-cost that marketing researchers seek and value in the direct single question approach.

The de-biasing procedure we have proposed shows considerable promise for improving marketing practice and we hope that the step we have taken will inspire more interest in improving single question approaches. Future research can test our de-biasing procedure in an even broader application context and may further illuminate some more subtle bias structures inherent in single question approaches.

In our investigation, we have looked into two particular lower priced consumer goods for which reference prices can be obtained. If anything, this setting renders any measured bias conservative as we believe that for such products the bias should be relatively smaller. For higher valued goods the bias tends to be higher (Murphy et al. 2005; Schmidt and Bijmolt 2019), and this may occur because they are less frequently purchased durable goods for which price preferences are remembered less compared to more frequently purchased non-durable goods (Estelami, Lehmann, Holden 2001). The bias may also be higher for more novel goods and for goods for which reference prices are harder to obtain because respondents may be more susceptible to biases induced by question formats when preferences are not well-developed yet and they are less familiar with the purchased item. Similarly, the magnitude of the bias may change along a product's life cycle. It may be higher early on and become smaller over the cycle. Finally, the bias may also vary across consumer types as it has been found that consumers who are low in purchase interest and low involvement tend to yield higher biases (Hofstetter et al. 2013). In sum, although the magnitude of the bias may vary across products and consumers, its structure should be rather consistent across contexts as it is introduced by the particular OE and DC question formats. Investigating the applicability of our de-biasing approach across various contexts may be a fruitful direction of future research. Finally, future research may compare the de-biasing approach proposed in this paper with alternative de-biasing approaches in the spirit of Miller et al. (2011), aiming at improving our understanding of when which approach will yield the most valid results.

# ***FIGURES***

FIGURE 1

ALTERNATIVE SINGLE QUESTION FORMATS

TO MEASURE CONSUMER WILLINGNESS TO PAY (WTP)

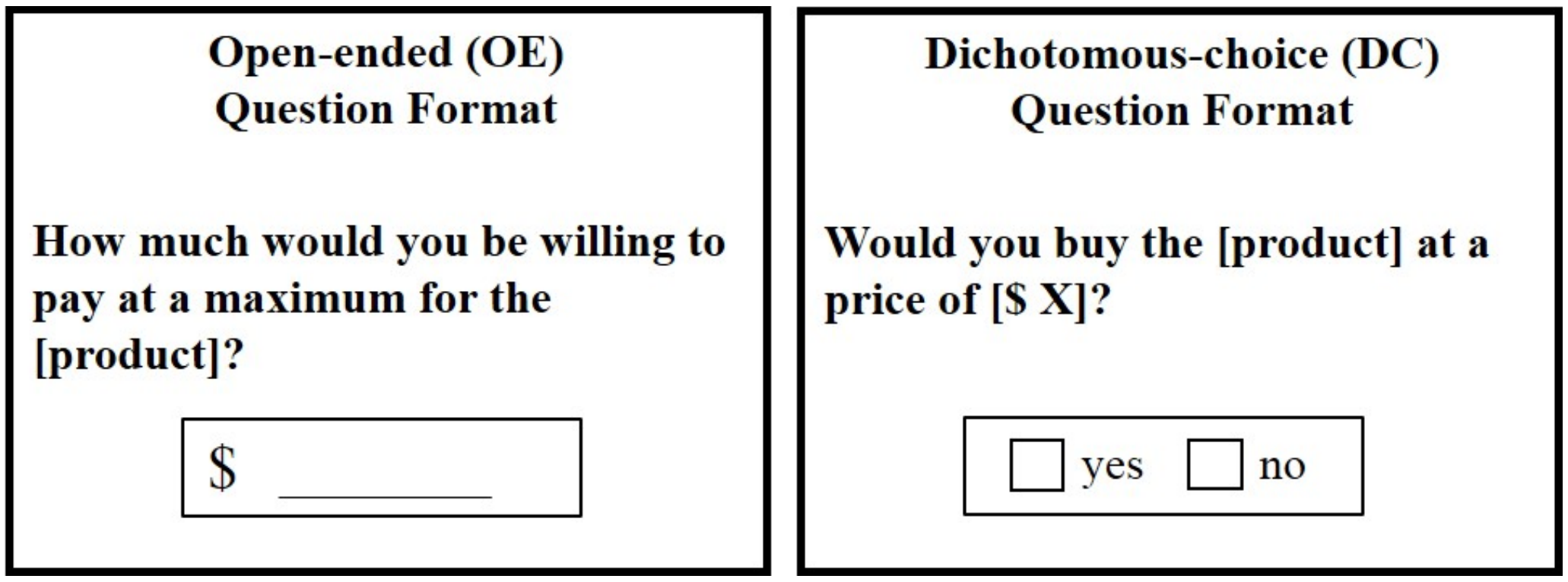

# FIGURE 2

## AGGREGATE DEMAND CURVES FOR UNIVERSITY GYM BAG BASED ON DIFFERENT WTP MEASUREMENT APPROACHES

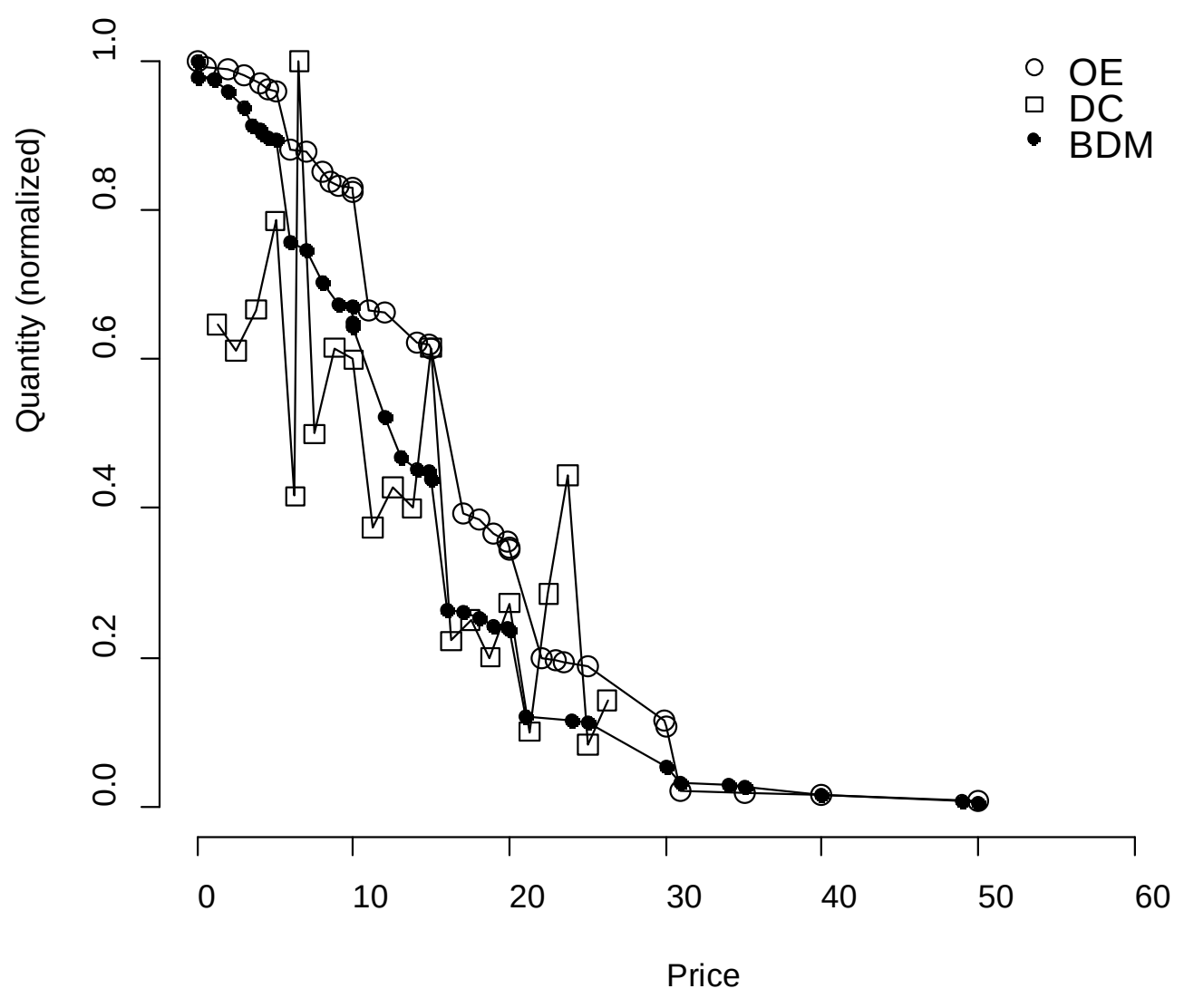

Note: The curves show the aggregate demand based on aggregating individuals' responses to the OE, DC, or BDM questions. OE: Open-ended question format; DC: Dichotomous-choice question format; BDM: Becker, DeGroot, and Marschak Mechanism; Quantity is the aggregate demand that is normalized to a range between 0 and 1 due to differing sample sizes. Prices are in CHF (Swiss Francs), which at the time of the study equal approximately to prices in USD.

# FIGURE 3

## SENSITIVITY ANALYSIS FOR DIFFERENCES IN PROFITS BY VARYING COV ($\boldsymbol{\theta}$, p)

**Study 1: University Gym Bag** | **Study 2: University Sweatshirt**

$cov(\theta, p)\epsilon[-3.08, 7.08]$ | $cov(\theta, p)\epsilon[-8.20, 1.80]$

*Difference in profits across different values of cov($\theta$, p) when $SD(\varepsilon_i) = 0$*

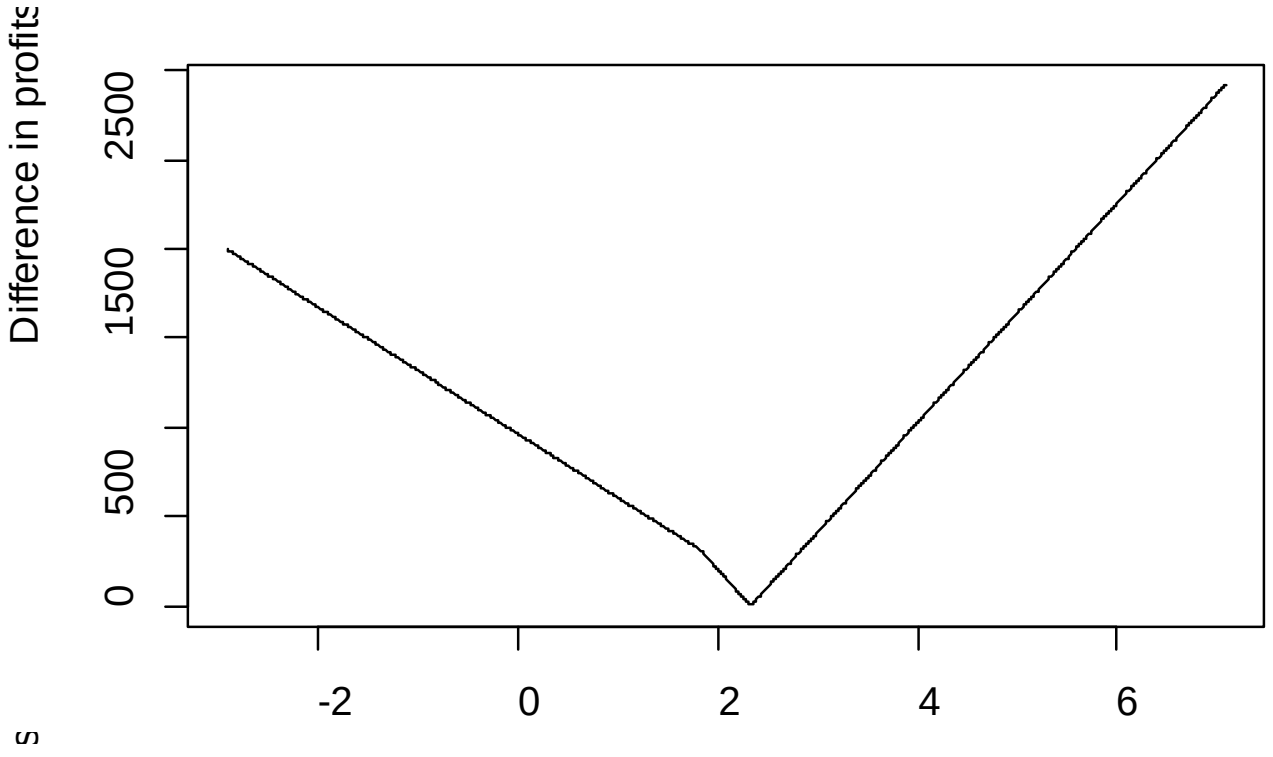

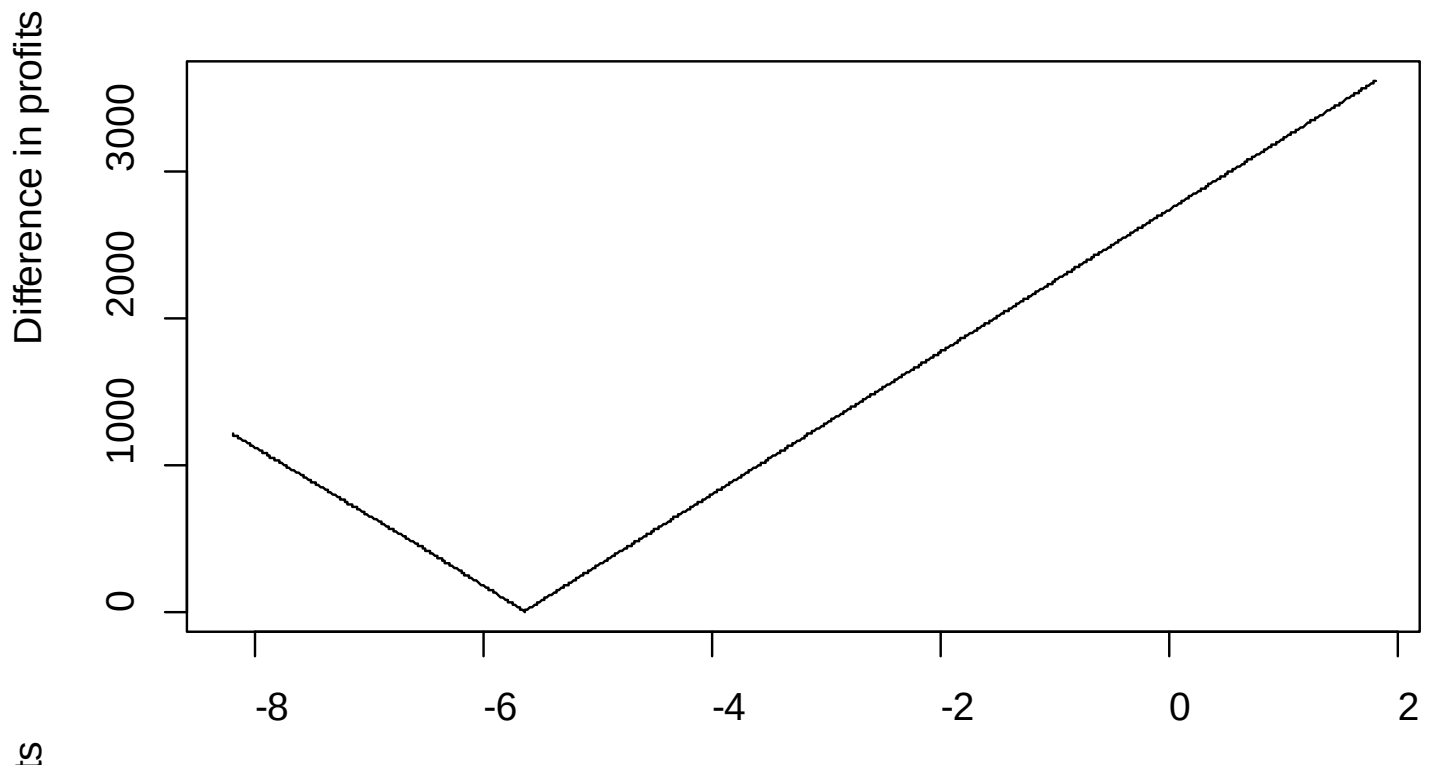

*Difference in profits across different values of cov($\theta$, p) when $SD(\varepsilon_i) = SD(\tilde{p}_i)$*

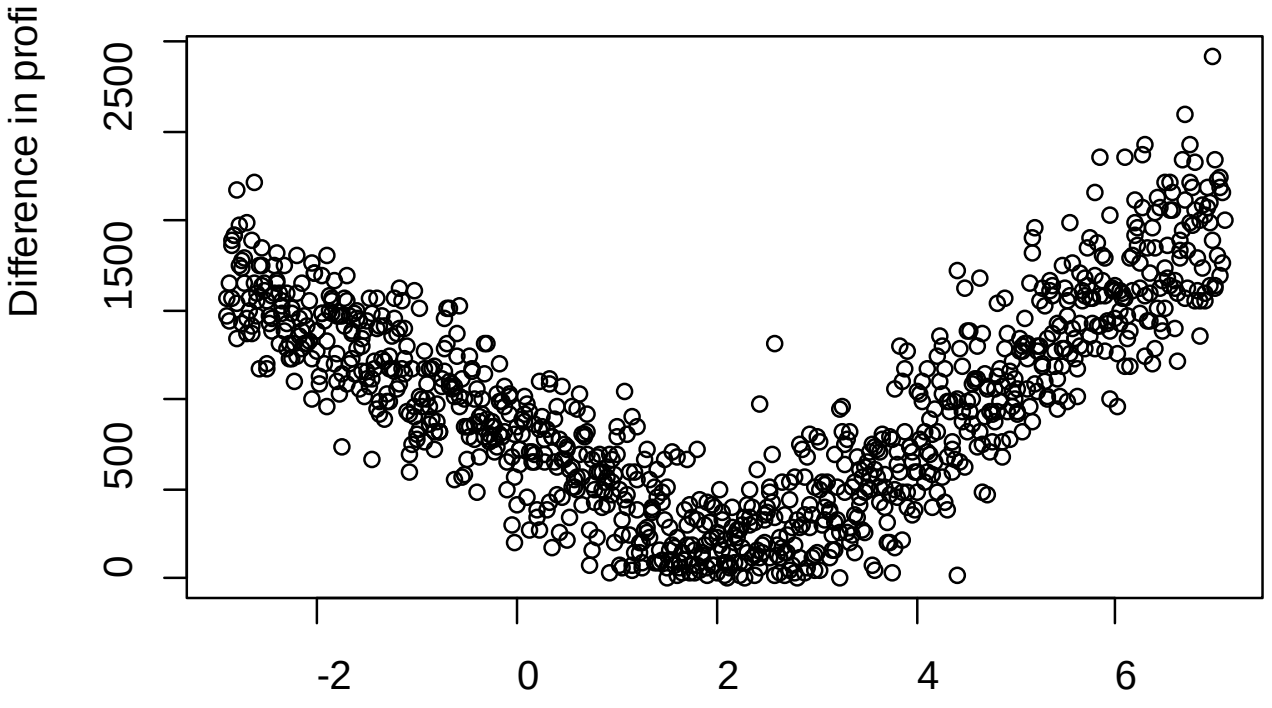

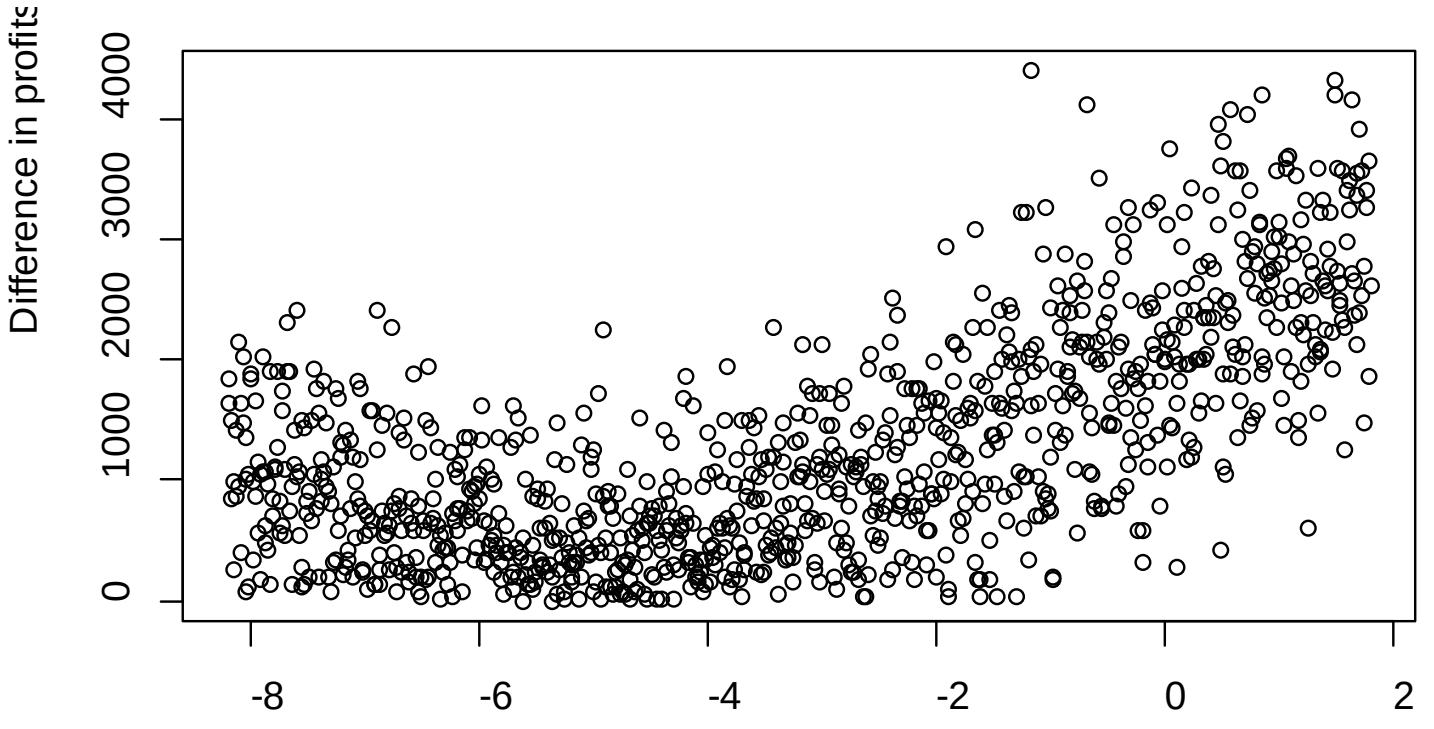

# *TABLES*

## TABLE 1

## ALTERNATIVE METHODS TO MEASURING CONSUMERS' WILLINGNESS TO PAY (WTP)

| | *Measurement* | | | |
|---|---|---|---|---|
| | *Direct* | | *Indirect* | |
| *Context* | *Single Question* | *Multiple Question* | *Single Question* | *Multiple Question* |
| *Hypothetical WTP (HWTP)* | e.g., Open-ended Question Format (OE) | e.g., Van-Westendorp Method (VWM) | e.g., Dichotomous-Choice Question Format (DC) | e.g., Choice-based Conjoint Analysis (CBC) |
| *Actual WTP (AWTP)* | e.g., Becker, DeGroot, and Marschak Mechanism (BDM) | e.g., Incentive-aligned Measurement of WTP Range (ICERANGE) | e.g., Incentive-aligned Dichotomous-Choice Question Format (IDC) | e.g., Incentive-aligned Choice-based Conjoint Analysis (ICBC) |
| *Hypothetical WTP (HWTP) De-Biasing Approaches* | e.g., Consequentialism, Honesty and Realism Approaches, Cheap Talk, Uncertainty Adjustment, Bayesian Truth Serum, Response Calibration | | e.g., Dual-response Choice Designs, Mental Simulation, Data Fusion of Hypothetical and Actual WTP Data | |

TABLE 2

COMPARISON OF ALTERNATIVE METHODS TO MEASURING CONSUMERS' WILLINGNESS TO PAY (WTP)

| **Measurement** | **Direct** | | | | **Indirect** | | | |
|---|---|---|---|---|---|---|---|---|
| Context | Hypothetical WTP | | Actual WTP | | Hypothetical WTP | | Actual WTP | |
| Example Methods | OE | VWM | BDM | ICERANGE | DC | CBC | IDC | ICBC |
| **Quality of Obtained Information** | | | | | | | | |
| Measure | Stated Preferences | | Revealed Preferences | | Stated Preferences | | Revealed Preferences | |
| Uncertainty on External Validity | High | | Low | | High | | Low | |
| Scope of Information | WTP of Entire Product | | WTP of Entire Product | | WTP of Entire Product and Product Features[a] | | WTP of Entire Product and Product Features[a] | |
| Capture Real Price Sensitivity | ++++ | | +++ | | ++ | | + | |
| **Ease of Implementation** | | | | | | | | |
| Ease of Data Collection | ++++ | | ++ | | +++ | | + | |
| Ease of Data Analysis | ++++ | | ++++ | | ++ | | ++ | |
| Interview Time | ++++ | | +++ | | ++ | | + | |
| Required Sample Size | ++++ | | +++ | | ++ | | + | |
| Costs | ++++ | | ++ | | +++ | | + | |
| **Applicability** | | | | | | | | |
| New Products | Yes | | No | | Yes | | No | |
| Existing Products | Yes | | Yes[b] | | Yes | | Yes[b] | |

Notes: ++++ Most favourable method; + Least favourable method. OE = Open-Ended (OE) Question Format. VWM = Van Westendrop Method. BDM = Becker, DeGroot, and Marschak Mechanism. ICERANGE = Incentive-aligned Measurement of WTP Range. DC = Dichotomous-Choice (DC) Question Format. CBC = Choice-based Conjoint Analysis. IDC = Incentive-aligned Dicotomous Choice Question Format. ICBC = Incentive-aligned Choice-Based Conjoint Analysis. [a] WTP for features of a product can only be obtained from indirect multiple question approaches such as choice-based conjoint analysis. We note that indirect single question approaches such as the dichotomous choice format or direct methods can be applied to individual features of a product, too. However, this would require a separate assessment of these individual features and subsequently more interview time as well as costs. [b] At minimum, incentive-aligned methods require an existing product prototype that can be sold to the survey respondents.

TABLE 3

MODEL NOTATION

| Notation | Meaning |
|---|---|
| $p_i$ | Actual WTP for consumer $i$ |
| $\tilde{p}_i$ | Stated WTP for consumer $i$ from an open-ended (OE) question format |
| $\hat{p}_i$ | Stated WTP for consumer $i$ from a dichotomous-choice (DC) question format |
| $p_i^c$ | Price cue presented to consumer $i$ in a dichotomous-choice (DC) question format |
| $\bar{p}$ | Mean of actual WTP $p_i$ |
| $\tilde{p}$ | Mean of stated WTP $\tilde{p}_i$ |
| $\hat{p}$ | Mean of stated WTP $\hat{p}_i$ |
| $p^c$ | Mean of price cues $p_i^c$ |
| $\alpha$ | Product category-level bias in an open-ended (OE) question format |
| $\varepsilon_i$ | Individual-level bias for consumer $i$ in an open-ended (OE) question format |
| $\theta_i$ | Individual-level bias for consumer $i$ in a dichotomous-choice (DC) question format |

# TABLE 4

## STATISTICAL ANALYSIS OF COLLECTED DATA AND DE-BIASED DATA OF UNIVERSITY GYM BAG DEMAND

| Collected Data | |
|---|---|
| *Data Source* | *Mean [Confidence Interval]* |
| OE | 16.046[a, b, c] [15.041, 17.051] |
| DC | 10.954[b] [10.402, 12.271] |
| BDM | 13.041 [12.037, 14.044] |
| **De-Biased Data** | |
| *Data Source* | *Mean [Confidence Interval]* |
| $BASIC_{[cov = 0, epsilon = 0]}$ | 11.072[a, b] [10.089, 12.055] |
| $EPSILON_{[cov = 0, epsilon = SD(OE)]}$ | 12.093[b] [10.851, 13.335] |
| $FULL_{[cov = 2.08, epsilon = SD(OE)]}$ | 13.854[b] [12.536, 15.172] |
| $FULL_{[cov = 2.33, epsilon = SD(OE)]}$ | 13.916[b] [12.580, 15.256] |
| BDM | 13.041 [12.037, 14.044] |

Note: OE: Open-ended question format; DC: Dichotomous-choice question format; BDM: Becker, DeGroot, Marschak Mechanism; BASIC, EPSILON, FULL: Refer to the steps of our de-biasing procedure; Values are shown with their 95% confidence interval in brackets; Superscript [a] indicates significant difference in terms of mean (t-test) relative to the benchmark; Superscript [b] indicates significant difference in terms of distribution (KS-test). For the DC data we used a Likelihood ratio test for the distribution comparison and compared confidence intervals (calculated based on Krinsky and Robb's (1986) procedure) for the mean comparison; Superscript [c] indicates non-overlapping confidence intervals relative to the BDM benchmark.

# TABLE 5

## ECONOMIC ANALYSIS RESULTS FOR UNIVERSITY GYM BAG

| *Data Source* | *Optimal Price* | *Optimal Quantity* | *Optimal Profit* | *Profit Percentage Difference to BDM* |
|---|---|---|---|---|
| OE | 15.000 [13.960, 15.990] | .615[c, d] [.546, .684] | 6,148.148[c, d] [5,557, 6,737] | 38.22%[c,d] |
| DC | 19.340[c, d] [15.500, 25.470] | .284[c, d] [.230, .337] | 4,024.250 [3,056, 5,315] | -9.52% |
| $BASIC_{[cov = 0, epsilon = 0]}$ | 14.810 [13.780, 15.830] | .360 [.270, .440] | 3,487.110[d] [2,916, 4,021] | -21.60%[d] |
| $EPSILON_{[cov = 0, epsilon =SD(OE)]}$ | 15.748 [12.590, 17.860] | .344 [.276, .427] | 3,702.102 [3,054, 4,260] | -16.77% |
| $FULL_{[cov = 2.08, epsilon = SD(OE)]}$ | 16.070 [11.470, 19.280] | .378 [.285, .489] | 4,182.080 [3,447, 4,709] | -5.97% |
| $FULL_{[cov = 2.33, epsilon = SD(OE)]}$ | 15.425 [7.910, 18.130] | .419 [.356, .609] | 4,362.976 [3,560, 4,929] | -1.91% |
| BDM | 14.900 [14.560, 15.190] | .449 [.389, .510] | 4,447.826 [3,865, 5,012] | N.A. |

Note: OE: Open-ended question format; DC: Dichotomous-choice question format; BASIC, EPSILON, FULL: Refer to the steps of our de-biasing procedure; BDM: Becker, DeGroot, Marschak Mechanism; Quantity scaled from [0, 1]; N.A. = not applicable; Values are shown with their 95% confidence interval in brackets. We checked for a non-overlapping of confidence intervals as a test for significant differences; Superscript [c] indicates non-overlapping confidence intervals; Superscript d indicates a significant difference at p = .05 (bootstrapping the difference between the measures); The shaded cells indicate that the confidence interval of the specific measure overlaps with the confidence interval of the corresponding benchmark measure obtained from our BDM data. Thus, shaded areas imply no statistical difference between the estimated measure and the benchmark.

# A De-biased Direct Question Approach to Measuring Consumers' Willingness to Pay

WEB APPENDIX - NOT FOR PUBLICATION

AUTHOR NAMES BLINDED FOR REVIEW

## *WEB APPENDIX A*

To better understand the adoption of demand measurement approaches in business practice, we conducted a brief survey among pricing managers in Switzerland. The basic population for our survey comprised all firms which are registered in the Swiss commercial registry. Based on a stratified sampling approach, which equally accounted for all sectors and firm sizes, we contacted 486 firms. After two weeks, we reminded all firms that had not yet participated to do so via the telephone. After this process, we ended up with 82 completed surveys, which corresponds to a completion rate of 17%. 65% of the completed surveys were answered by members of the senior management, 22% by department heads of marketing, and 13% by other functions within the firm. With regard to economic sectors, 44% of the firms covered in our survey came from industry, 39% from services, and 16% were in retailing. Firm revenue was almost evenly distributed and ranged from less than CHF 1 Mio. to more than CHF 500 Mio. Similarly, market shares ranged from less than 1% to more than 50% in our sample. Finally, the number of competitors was evenly distributed and ranged from monopoly, state-owned firms to firms in highly-competitive markets with more than 50 major competitors. We administered a survey consisting of 53 questions regarding pricing-related topics to the firms. Here, we report the results of the question that referred to the adoption of demand measurement approaches in business practice (see Figure A.1).

FIGURE A.1

ADOPTION OF DEMAND MEASUREMENT APPROACHES IN BUSINESS PRACTICE

Which demand measurement approaches are you using in your firm?

| **Demand Measurement Approaches** | **absolute** | **in percent*** |
|---|---|---|
| Direct Approach | **23** | **28%** |
| Analysis of Market Data | **20** | **24%** |
| Field Experiments | 3 | 4% |
| Lab Experiments | 1 | 0% |
| Conjoint Analysis | 3 | 4% |
| Experimental Auctions | 0 | 0% |
| Experimental Lotteries | 0 | 0% |
| Expert Judgements | 3 | 4% |
| Other | 3 | 4% |

* Notes: N = 82 respondents. Multiple responses possible.

***WEB APPENDIX B***

We assume $\theta_i \sim f(\,)$ with mean$(\theta_i) = 0$. What we say is

1. f( ) need not be symmetrical
2. we can still have bias if f( ) is not symmetrical and mean $(\theta_i) = 0$ ,

To show this:

1. Distribution function: $x_1 y_1 + x_2 y_2 = 1$ where $x_1$ is the range below zero with density $y_1$ and $x_2$ the range above zero with density $y_2$.
2. Mean = 0

Mean $= \int_{-x_2}^{0} x\, y_2 dx + \int_{0}^{x_1} x\, y_1 dx$

$= y_2 [\frac{1}{2} x^2]_{-x_2}^{0} + y_1 [\frac{1}{2} x^2]_{0}^{x_1}$

$= -y_2 x_2^2 \frac{1}{2} + y_1 \frac{1}{2} x_1^2$

Set mean = 0

1. and 3. can exist

$y_1 x_1^2 - y_2 x_2^2 = 0$ 3.

Proof: Let

$x_1 y_1 = 0.7$

$x_2 y_2 = 0.3$

$x_2 = \frac{0.7}{0.3} = \frac{7}{3}$

$0.7\, x_1 - 0.3 x_2 = 0$

$0.7\, x_1 = 0.3 x_2$

$x_1 = \frac{3}{7} x_2$

Specifically:

$x_2 = 7$

$x_1 = 3$

$y_2 = \frac{0.3}{7}$

$y_1 = \frac{0.7}{3}$

If

$x_1 = 1$

$x_2 = \frac{7}{3}$

$y_1 = 0.7$

$y_2 = 0.3 \frac{3}{7} = \frac{0.9}{7}$

Basically, this shows that with mean $(\theta)$ = 0, you can still have lots of biases.

***WEB APPENDIX C***

$$
\begin{aligned}
cov(\theta, p) &= \frac{1}{n} \textstyle\sum_{i=1}^{n}(\theta_i - \bar{\theta})(p_i - \bar{p}) \\
&= \frac{1}{n} \textstyle\sum_{i=1}^{n}[(\theta_i p_i) - \theta_i \bar{p}] \\
&= \frac{1}{n} \textstyle\sum_{i=1}^{n} \theta_i \, p_i - \frac{1}{n} \textstyle\sum_{i=1}^{n} \theta_i \, \bar{p} \\
&= \frac{1}{n} \textstyle\sum_{i=1}^{n} \theta_i \, p_i - \frac{\bar{p}}{n} \textstyle\sum_{i=1}^{n} \theta_i \quad \text{as} \quad \frac{\bar{p}}{n} \textstyle\sum_{i=1}^{n} \theta_i = 0 \\
&= \frac{1}{n} \textstyle\sum_{i=1}^{n} \theta_i \, p_i
\end{aligned}
$$

## *WEB APPENDIX D*

### *STIMULUS STUDY 1: UNIVERSITY GYM BAG*

Below we include the stimulus used in Study 1. We blinded the university logo for review and will unblind them if our paper is accepted for publication.

| | |
|---|---|
| 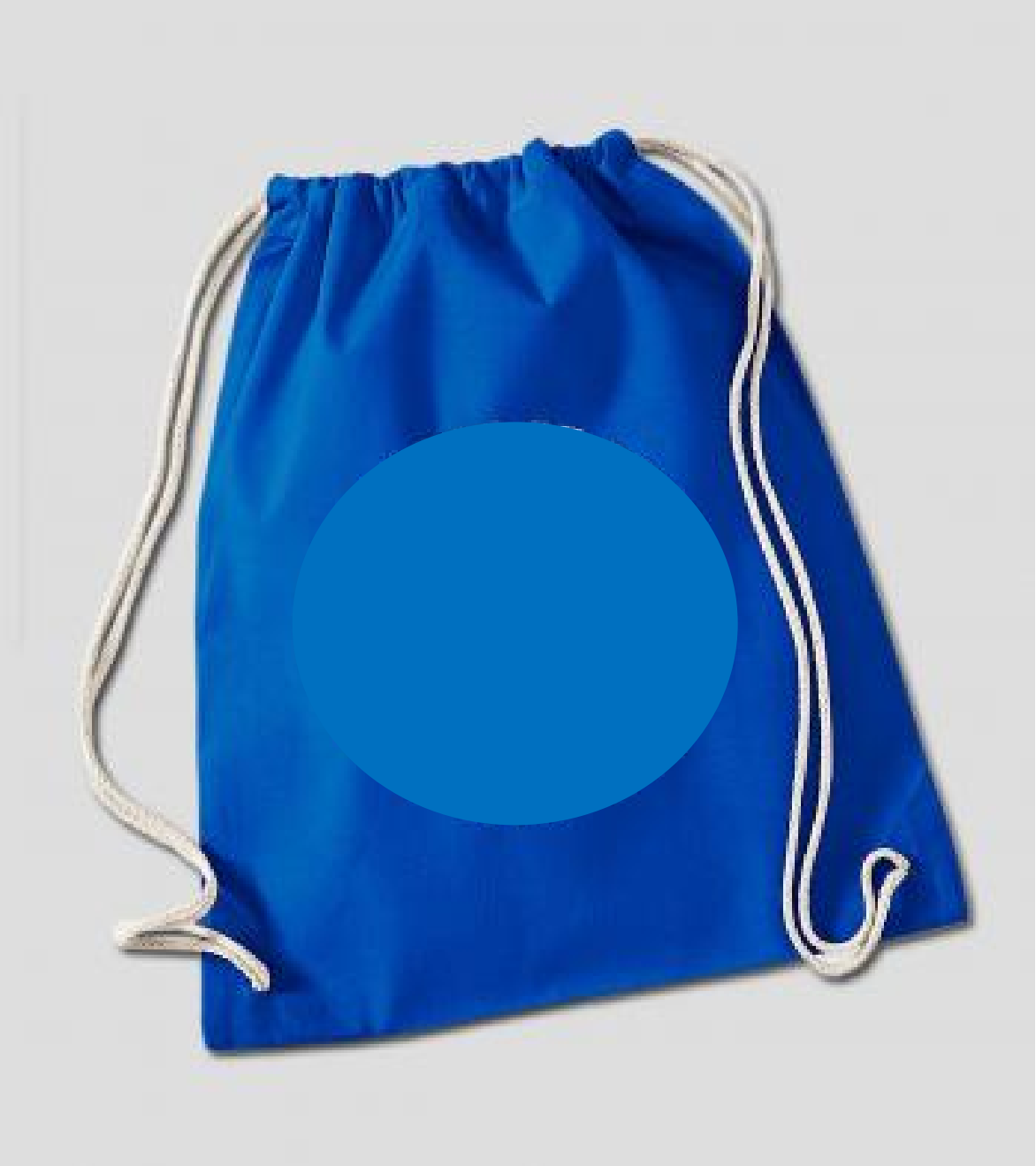 | ▪ Functional university gym bag.<br>▪ Ideal for party, shopping, university, and sports.<br>▪ 100% organic cotton.<br>▪ High quality screen print.<br>▪ Printed in Switzerland by hand.<br>▪ Color: royal blue.<br>▪ Size: 37 x 46 cm.<br>▪ Free shipping. |

## *WEB APPENDIX E*

## *TRANSPARENCY AND ACCEPTABILITY RATING QUESTIONS*

Note: The specific experimental groups are denoted as OE (Open-ended Question Format), DC (Dichotmous-choice Question Format), and BDM (Becker, deGroot, Marschak Mechanism).

1. If the gym bag (sweatshirt) was actually offered to you for purchase in an online shop, how certain are you that you would purchase the product at **your** (OE/BDM) / **the** (DC) stated price? (1 = “very uncertain,” 7 = “very certain”)
2. Is it clear to you why it is in your best interest to **state** (OE/BDM) / **accept** (DC) exactly the price you are willing to pay for the gym bag (sweatshirt)? (1 = “not at all,” 7 = “very much so”)
3. OE/DC: Was it confusing to you to state your maximum willingness to pay for the gym bag (sweatshirt)? (1 = “not at all,” 7 = “very much so”); BDM: Was the buying process confusing to you? (1 = “not at all,” 7 = “very much so”)
4. BDM only: Did you understand the buying process? (1 = “not at all,” 7 = “very much so”)
5. OE/DC/BDM: This task was very easy to understand and complete. (1 = “not at all,” 7 = “very much so”)
1. OE/DC/BDM: Did you perceive this task as fair? (1 = “not at all,” 7 = “very much so”)
2. OE/DC/BDM: I will be happy to do this task again in the future. (1 = “not at all,” 7 = “very much so”)

# *WEB APPENDIX F*

FIGURE F.1
AGGREGATE DEMAND CURVES FOR UNIVERSITY GYM BAG
BASED ON DIFFERENT WTP MEASUREMENT AND DE-BIASING APPROACHES

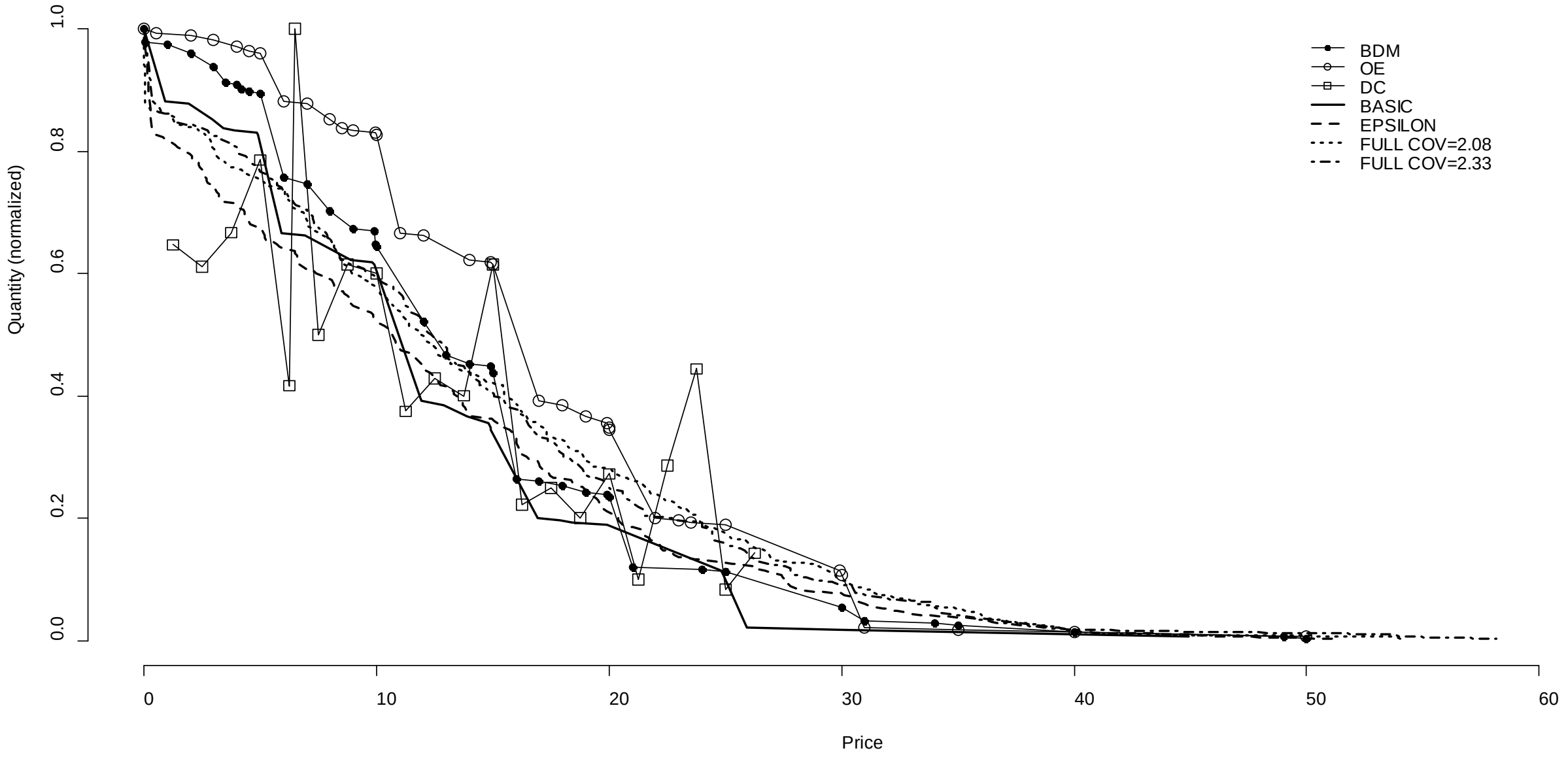

Note: The curves show the aggregate demand based on aggregating individuals' responses to the OE, DC, or BDM questions and based on the different de-biasing approaches. OE: Open-ended question format; DC: Dichotomous-choice question format; BDM: Becker, DeGroot, and Marschak Mechanism; Quantity is the aggregate demand that is normalized to a range between 0 and 1 due to differing sample sizes. Prices are in CHF (Swiss Francs), which at the time of the study equal approximately to prices in USD.

## *WEB APPENDIX G*

## *UNIVERSITY SWEATSHIRT STUDY (STUDY 2)*

We replicated our de-biasing procedure demonstrated in the university gym bag study (Study 1) in a second study using a university sweatshirt as a stimulus.

### *METHOD AND DATA COLLECTION*

***Participants***. The data for the university sweatshirt study was also collected through an online experiment. To recruit participants, we sent out 12,448 invitation emails to the entire student body (undergraduate, graduate, and Ph.D. candidates) of a large Swiss university. We motivated participation by offering all survey participants a chance to win an Apple iPhone 7 Plus in a raffle[1]. The participants were further informed that their chance to win the raffle was independent of their experimental responses[2]. A total of 772 participants chose to take part in the sweatshirt-study, which represents a response rate of 6.18%. We pre-specified that data collection would end after seven days (i.e., the decision to stop collecting data was independent from the experimental results; we did not analyze the data until after data collection for the given study had been completed). Within the seven-day period, we collected as much data as we could.

***Stimulus.*** The stimulus we used in our second study was a hooded sweatshirt imprinted with a logo of the university that was not available in the market at the time of the study (see Figure G.1 for a depiction of the stimulus). The sweatshirt was designed and fabricated exclusively for the purpose of the study. We expected the university sweatshirt to be both interesting and affordable for most of the students, who actually represent the target segment of the product. As apparel and accessories are also often sold online, the online channel represents the known and accustomed distribution channel for these categories.[3]. Further, because the students had no reference for the exact market price for this distinctive university sweatshirt, their HWTP or AWTP statements would not be capped[4].

As our stimulus were new to the market, no repeat purchases were observed and participants were allowed to state their WTP only once. Further, the university sweatshirt was not displayed in a competitive setting. This meant that students were unable to select the stimulus from a group of competing products as they would be able to in a real online store. Finally, because we were using an online experiment, shipping the product had to be easy and cheap, which was the case with the sweatshirt.

---

[1] We used the smartphone as a single incentive to motivate participation in our survey in order to recruit an adequate number of subjects. However, the smartphone was not connected to our stimulus and the incentive-aligned condition under BDM, where proper incentives are offered to the participants so that they are motivated to reveal their true preferences (see Wertenbroch and Skiera 2002 for details).

[2] It is possible that some consumers may have taken part in the survey just to win the smartphone and may not have been interested in purchasing the hooded sweatshirt. However, there is no reason to believe that they have affected our statistical analyses as subjects were randomly assigned to different experimental groups.

[3] See e.g., the Stanford Bookstore online: https://www.bkstr.com/stanfordstore/home.

[4] We acknowledge, however, that some participants may have a reference price from similar products in mind.

FIGURE G.1
STIMULUS STUDY 2: UNIVERSITY SWEATSHIRT

Below we include the stimulus used in Study 2. (We blinded the university logo for review and will unblind them in case our paper gets accepted for publication.)

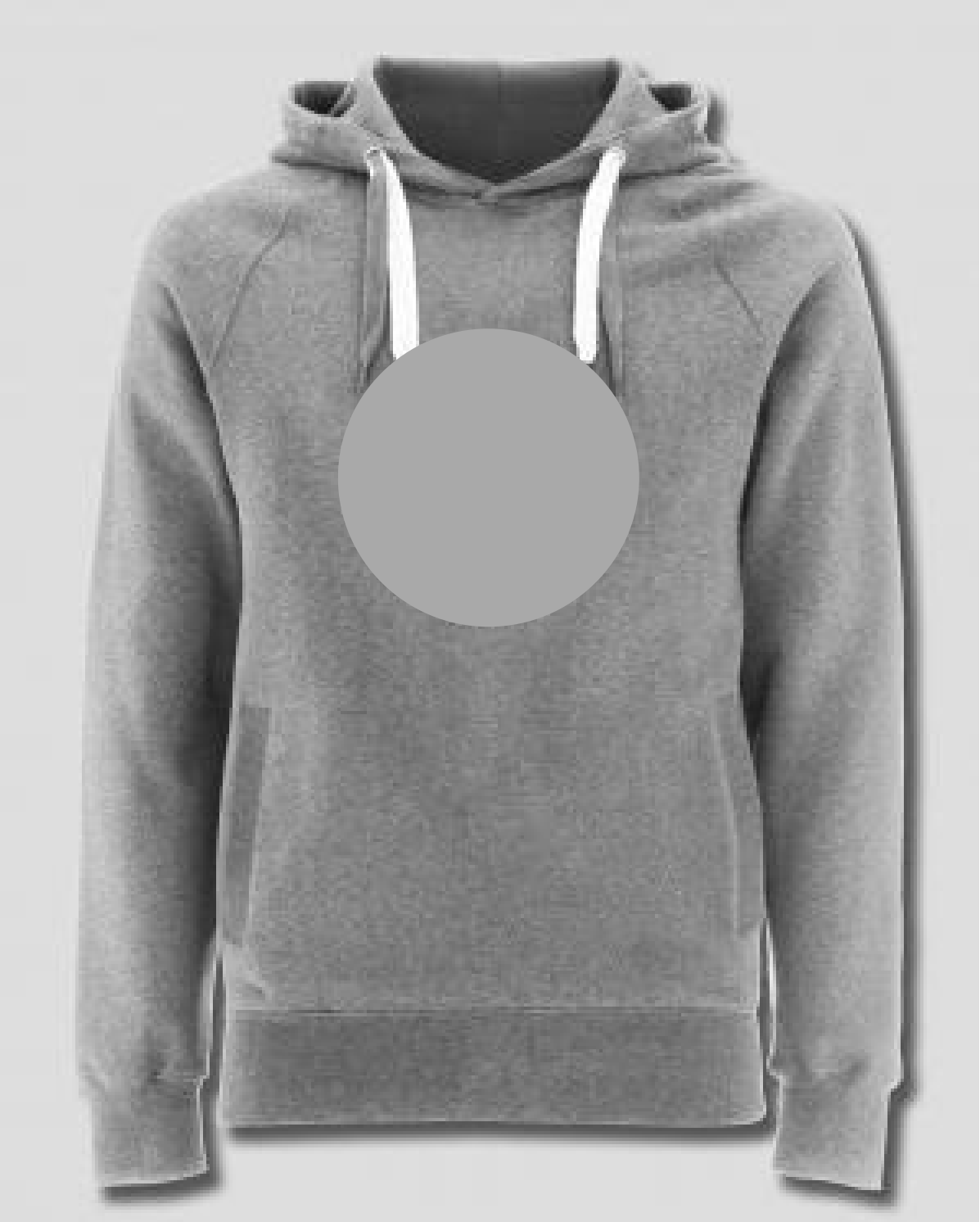

- Design and ethics combined in a high-quality university sweatshirt (unisex).
- Fair production according to high environmental and social standards.
- Certification according to Oekotex Standard 100.
- 100% organic cotton.
- Durable screen printing.
- Printed in Switzerland by hand.
- Available in all sizes from S to XL.
- Free shipping.

***Experimental Design.*** We developed three different independent experimental groups per study and used a between-subjects design. Each participant was randomly assigned to one treatment group.

In the open-ended (OE) question format group, each participant directly stated her hypothetical WTP for the university sweatshirt.

In the dichotomous-choice (DC) question format group, we used a total number of 21 price levels, ranging from CHF 10 up to CHF 110 incremented in steps of CHF 5 for the sweatshirt study. We set the market price because the most expensive sweatshirt similar to our products as our upper limit was CHF 110. Each respondent received a price level that was chosen randomly out of the 21 available levels. The random distribution of the price levels was even, meaning that all the price levels had an equal chance of one in 21 to appear in the respondent's DC question.

For the BDM group, our control group for validating our de-biasing procedures, we determined our benchmarking AWTP data by using an incentive-aligned mechanism, the BDM mechanism proposed by Becker, DeGroot, and Marschak (1964). We chose the BDM mechanism as WTP from BDM has been found to not significantly differ from WTP based on real purchase data (Miller et al. 2011). We implemented the BDM mechanism in a way similar to what Wertenbroch and Skiera (2002) did. In our particular application of the BDM mechanism, we told participants that they would have a chance to purchase the university sweatshirt without having to invest more money than they would be willing to pay for the product, that the price for the university sweatshirt was not yet set, and that it would be determined randomly from a predefined uniform distribution unknown to the participants. Participants were further told that they were obligated to buy the university sweatshirt at the randomly determined price if the price was less than or equal to their stated WTP. However, if the randomly determined price was higher, a respondent would not have a chance to buy the product. This mechanism ensures that participants have no incentive to state a price that is higher or lower than their true WTP.

To carry out the buying obligation, we recorded the name and address of each participant in the BDM group. After the completion of the study, we determined each individual participant's buying obligation by drawing from a discrete uniform distribution which corresponded to the price-levels used under the DC question format. The distribution thus ranged from CHF 10 to CHF 110 incremented in steps of CHF 5 and included a total of 21 price levels. We determined per participant whether the randomly drawn price was smaller or equal to her stated WTP. Thus, none of the participants had to purchase the sweatshirt at a price that was larger than their stated WTP. Out of all participants in the BDM group, 12.70% were obliged to buy the product at an average price of 26.77 (SD = 16.26, min = 10, max = 75). Only 5 participants (16.13%) of those who were obliged to buy paid a price higher than the average BDM WTP of 37.12. After the completion of the study, all participants who were obligated to buy were sent the sweatshirt with an invoice, via the post. The invoice was due within 14 days and payable with cash or credit card. (This payment process was officially approved by the appropriate university authorities). In the sweatshirt study, out of all 244 respondents in the BDM group, 31 (12.7%) were required to purchase the sweatshirt[5]. Only one respondent refused to comply with her purchase obligation and returned the product[6].

We obtained 262 responses in the OE group, 266 responses in the DC group, and 244 responses in the BDM group. Our realized sample size exceeds current expectations in experimental studies of larger than 50 respondents per cell (Simmons 2014).

The three experimental groups did not differ significantly in terms of socio-demographics or socioeconomic status. We performed a multivariate analysis of variance for the sweatshirt-study (Pillai-Spur: F = .984, p = .480) for age (F = .925, p = .428), sex (F = .879, p = .452), education (F = 1.696, p = .167), occupation (F = 1.712, p = .164), income (F = .071, p = .975), budget for clothing and accessories (F = .735, p = .532), and purchase interest (F = .695, p = .555).

***Experimental Procedure***. We divided our online experiment into three parts. The first part described the product (i.e., the university sweatshirt) in the OE, DC, and BDM groups. The second part consisted of the WTP task in the different experimental treatment groups. In the third part of the online experiment, we conducted a brief survey on the respondents' socio-demographics and economics, and we made sure the participants understood the WTP elicitation method to which they were exposed (see Web Appendix D of the main paper for further details).

---

[5] In the sweatshirt study, we removed one outlier with a BDM value of CHF 200 who explicitly stated that she intentionally and falsely over-reported her true WTP.

[6] Valid AWTP estimation requires that the respondents understand the BDM procedure and the underlying buying process. In our sample, respondents understood the BDM mechanism quite well. As Wertenbroch and Skiera (2002) and Miller et al. (2011) did, we asked the subjects if it was clear to them why it was in their best interest to state exactly the price they were willing to pay. Using a seven-point Likert scale from one (not at all) to seven (very much so), the participants responded with an average of 5.803 in the sweatshirt-study. We used a similar method to determine the understanding of the buying process and found an average of 6.070 for the sweatshirt-study. Finally, we asked respondents if stating their WTP for the product was a task which was easy to understand and complete and participants replied with an average of 6.489 in the sweatshirt-study (see Web Appendix E for the exact wording).

***WTP Estimation Procedure.*** Figure G.2 plots the observed demand in each treatment group. For the OE and BDM groups, we obtained respondent's hypothetical WTP and actual WTP directly from the survey data and plot demand $q(p)$ as the probability that a respondent`s WTP is equal to or greater than a certain price $p$ using the demand function of the form $q(p) = Pr(WTP \geq p)$. For the DC group we plotted the choice share for each price level. We determine the face validity of WTP measures by correlating elicited WTP with the respondent`s purchase interest. We measured purchase interest itself by using a seven-point Likert scale ranging from one (low interest in the product) to seven (high interest in the product). Face validity is high for all methods because correlations are positive and significant (OE: r = .379, p = .000, BDM: r = .314, p = .000). We did not test the DC question format because HWTP data is not available on an individual level here.

## *RESULTS*

The university gym bag study (Study 1) demonstrated the great promise of our de-biasing procedure for the direct single question approach. In this study, we will apply the same procedure to a different data set for a university sweatshirt we have collected to investigate the robustness of our de-biasing approach.

As shown in G.2, the demand functions show a similar pattern compared to the gym bag data. At first glance, the OE demand is consistently inflated, while the DC demand appears to be rotated counter-clockwise. A closer inspection also confirms, as shown in Table F.1, that the OE data series is significantly different from the BDM both in mean [$t(543.92)_{\text{OE vs. BDM}} = 4.166$, $p < .01$] and distribution ($D_{\text{KS-Test, OE vs. BDM}} = .181$, $p < .01$; $D_{\text{LR-Test, OE vs. BDM}} = 38.65$, $p < .01$). In addition, the 95%-confidence intervals for the OE mean and BDM mean do not overlap. As expected, the OE mean is significantly inflated relative to the BDM. However, the DC data series is closer to the BDM and passes all tests except the LR-test ($D_{\text{LR-Test, DC vs. BDM}} = 17.681$, $p < .01$).

Table D.2 also confirms that our partial, BASIC and EPSILON, as well as FULL de-biasing procedures all have improved on the OE data series. All three de-biased data series pass all the tests except the KS-test, similar to the DC data series [$t(502.67)_{\text{BASIC vs. BDM}} = 1.86$, $p > .05$, $t(476.76)_{\text{EPSILON vs. BDM}} = 1.22$, $p > .22$, $t(472.05)_{\text{FULL (cov = -3.20) vs. BDM}} = .49$, $p > .05$, $t(475.06)_{\text{FULL (cov = -5.64) vs. BDM}} = -.96$, $p > .05$], ($D_{\text{KS-Test, BASIC vs. BDM}} = .17$, $p < .01$, $D_{\text{KS-Test, EPSILON vs. BDM}} = .14$, $p < .05$, $D_{\text{KS-Test, FULL (cov = -3.20) vs. BDM}} = .14$, $p < .05$, $D_{\text{KS-Test, FULL (cov = -5.64) vs. BDM}} = .20$, $p < .01$). However, the estimated means from our de-biased data series are significantly closer to the BDM mean and two of them beat the DC mean.

FIGURE G.2
AGGREGATE DEMAND FOR UNIVERSITY SWEATSHIRT
BASED ON DIFFERENT WTP MEASUREMENT APPROACHES

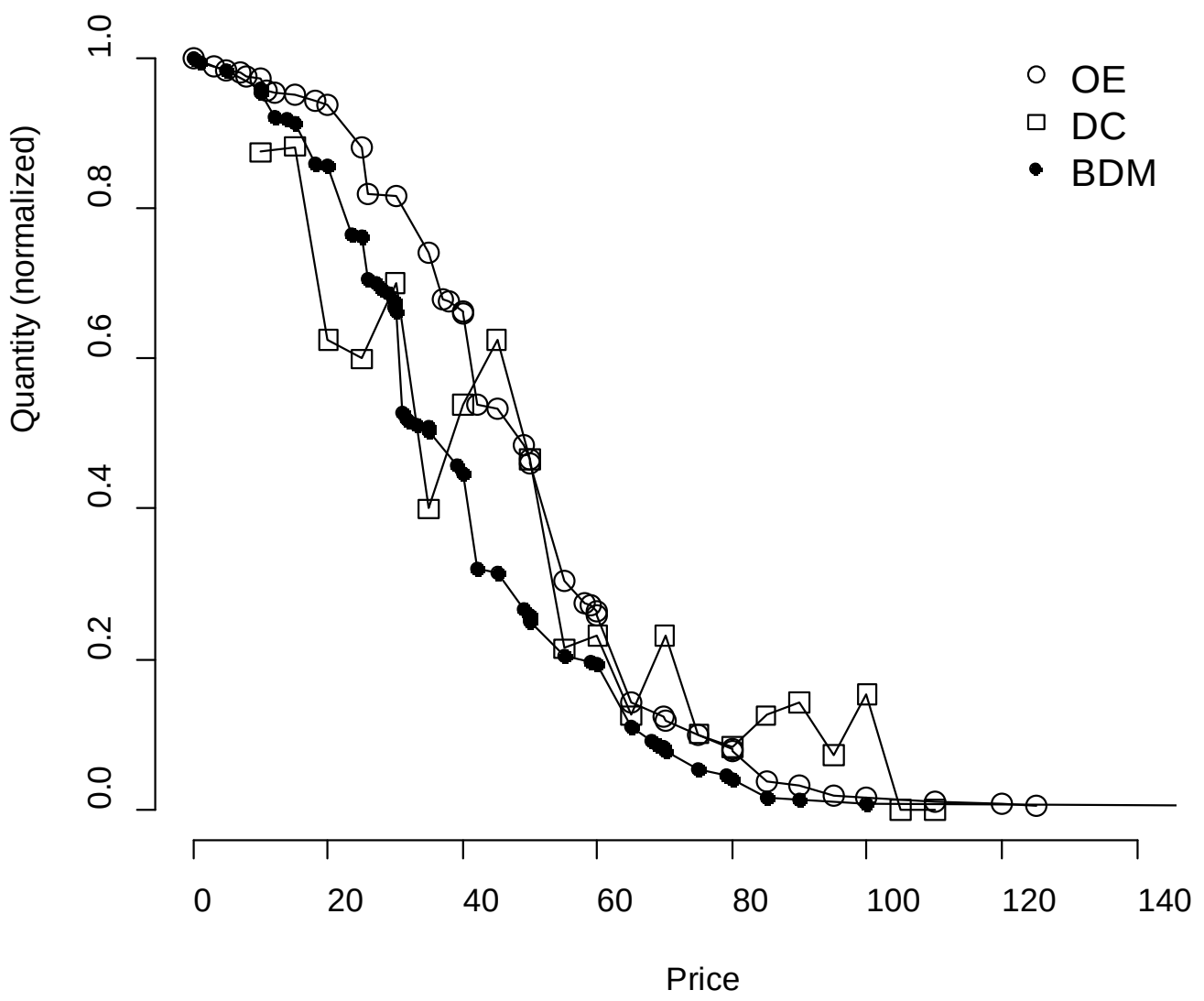

Note: The curves show the aggregate demand based on aggregating individuals' responses to the OE, DC, or BDM questions. OE: Open-ended question format; DC: Dichotomous-choice question format; BDM: Becker, DeGroot, and Marschak Mechanism; Quantity is the aggregate demand that is normalized to a range between 0 and 1 due to differing sample sizes. Prices are in CHF (Swiss Francs), which at the time of the study equal approximately to prices in USD.

To see how these improvements might affect managerial decisions, we conducted a similar economic analysis as in Study 1. Here, we used marginal costs for the university sweatshirt as obtained from the manufacturer of c = CHF 15.00. Table F.2 summarizes the results from this analysis. Similar to the previous study, the OE data series on the university sweatshirt generated statistically different optimal quantity and the DC data series different optimal price estimates. More importantly, both data series led to wildly overestimated profits, 48.83% and 23.03% respectively. The first BASIC de-biasing procedure subtracting only $\tilde{p} - \hat{p}$ improved significantly on the original OE data, but did not beat the DC data series. The second one, EPSILON, where individual-specific variation is accounted for, improved on DC significantly and the profit estimate differed from BDM by only 6.16%. As Table E.3 further shows, our FULL de-biasing procedure performs, once again, remarkably well. The estimates of the optimal price and quantity are not statistically different from BDM and the estimate of optimal profits is within 5.26%, -.01% of the BDM. Please see Figure G.3 for a visualization of the different demand curves resulting from these de-biasing approaches.

# TABLE G.1
# STATISTICAL ANALYSIS OF COLLECTED DATA AND DE-BIASED DATA OF UNIVERSITY SWEATSHIRT

| Collected Data | |
|---|---|
| *Data Source* | *Mean [Confidence Interval]* |
| OE | 45.758[a, b, c] [43.277, 48.250] |
| DC | 40.324[b] [34.878, 45.550] |
| BDM | 37.120 [34.677, 39.564] |
| **De-Biased Data** | |
| *Data Source* | *Mean [Confidence Interval]* |
| BASIC$_{[cov = 0, epsilon = 0]}$ | 40.40[b] [37.92, 42.87] |
| EPSILON$_{[cov = 0, epsilon = SD(OE)]}$ | 39.63[b] [36.40, 42.86] |
| FULL$_{[cov = -3.20, epsilon = SD(OE)]}$ | 38.81[b] [35.69, 41.93] |
| FULL$_{[cov = -5.64, epsilon = SD(OE)]}$ | 35.14[b] [31.89, 38.40] |
| BDM | 37.12 [34.68, 39.56] |

Note: OE: Open-ended question format; DC: Dichotomous-choice question format; BDM: Becker, DeGroot, Marschak Mechanism; BASIC, EPSILON, FULL: Refer to the steps of our de-biasing procedure; Values are shown with their 95% confidence interval in brackets; Superscript [a] indicates significant difference in terms of mean (t-test) relative to the benchmark; Superscript [b] indicates significant difference in terms of distribution (KS-test). For the DC data we used a Likelihood ratio test for the distribution comparison and compared confidence intervals (calculated based on Krinsky and Robb's (1986) procedure) for the mean comparison; Superscript [c] indicates non-overlapping confidence intervals relative to the BDM benchmark.

# TABLE G.2
# ECONOMIC ANALYSIS RESULTS FOR UNIVERSITY SWEATSHIRT

| *Data Source* | *Optimal Price* | *Optimal Quantity* | *Optimal Profit* | *Profit Percentage Difference to BDM* |
|---|---|---|---|---|
| OE | 39.9 [26.330, 44.430] | .664[c,d] [.574, .911] | 16,536[c,d] [14540, 17,947] | 48.83%[c,d] |
| DC | 49.112[d] [43.354, 55.342] | .402 [.335, .4765] | 13,670[d] [11,173, 16,2701] | 23.03%[c,d] |
| BASIC [cov = 0, epsilon = 0] | 43.566 [39.100, 48.780] | .484 [.364, .579] | 13,846[d] [12,041, 15,460] | 24.62%[d] |
| Epsilon [cov = 0, epsilon = SD(OE)] | 43.614 [34.990, 54.630] | .412 [.224, .512] | 11,795 [9,947, 13,020] | 6.16% |
| FULL [cov = -3.20, epsilon = SD(OE)] | 44.748 [36.360, 54.360] | .393 [.249, .490] | 11,695 [9,834, 13,215] | 5.26% |
| FULL [cov = -5.64, epsilon = SD(OE)] | 43.266 [37.580, 46.550] | .389 [.328, .477] | 11,004 [9,148, 12,603] | -.01% |
| BDM | 40 [35.520, 45.510] | .444 [.331, .530] | 11,111 [9,605, 12,573] | N.A. |

Note: OE: Open-ended question format; DC: Dichotomous-choice question format; BASIC, EPSILON, FULL: Refer to the steps of our de-biasing procedure; BDM: Becker, DeGroot, Marschak Mechanism; Quantity scaled from [0, 1]. N.A. = not applicable; Values are shown with their 95% confidence interval in brackets; We checked for a non-overlapping of confidence intervals as a test for significant differences; Superscript [c] indicates non-overlapping confidence intervals; Superscript [d] indicates significant difference at p = .05 (bootstrapping the difference between the measures); The shaded cells indicate that the confidence interval of the specific measure overlaps with the confidence interval of the corresponding benchmark measure obtained from our BDM data. Thus, shaded areas imply no statistical difference between the estimated measure and the benchmark.

# FIGURE G.3
# AGGREGATE DEMAND FOR UNIVERSITY SWEATSHIRT
# BASED ON DIFFERENT WTP MEASUREMENT AND DE-BIASING APPROACHES

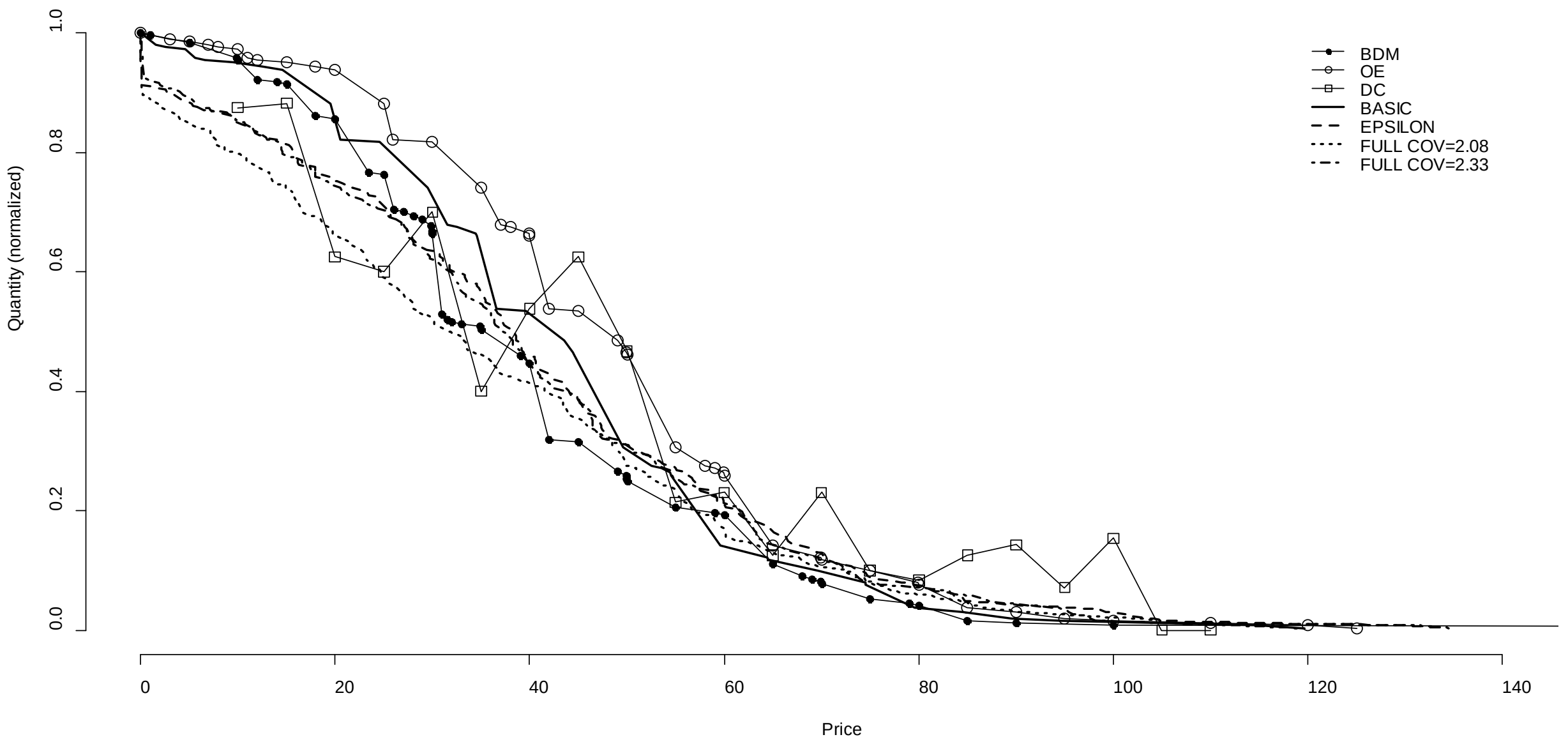

Note: The curves show the aggregate demand based on aggregating individuals' responses to the OE, DC, or BDM questions and based on the different de-biasing approaches. OE: Open-ended question format; DC: Dichotomous-choice question format; BDM: Becker, DeGroot, and Marschak Mechanism; Quantity is the aggregate demand that is normalized to a range between 0 and 1 due to differing sample sizes. Prices are in CHF (Swiss Francs), which at the time of the study equal approximately to prices in USD.

## *WEB APPENDIX H*

## *ROBUSTNESS OF RESULTS TO THE DC PRICE RANGE*

***Setup of Simulation.*** We explored the sensitivity of our results to the chosen DC price range in a simulation study. We followed the setup of our experimental design, but generated WTP values for OE, DC, and BDM from a known distribution. This allowed us to know a consumer's actual WTP and to keep the number of individuals in the DC group constant, despite narrowing the range of DC price levels (i.e., by removing price levels)[7].

We randomly drew WTP values from a normal distribution with a mean of CHF 50 and standard deviation of CHF 10 within a range between CHF 15 and CHF 85. We randomly drew 250 actual WTP values in each of the three experimental groups, OE, DC, and BDM. We biased the randomly drawn OE and DC data according to our bias models. For OE, we set the product category-specific inflator $\alpha = \frac{1}{2}\,\tilde{p}$ (i.e., half the mean of the OE data from the university sweatshirt study; see Web Appendix F). We sampled $\theta_i$ randomly at the individual-level according to the function described in Web Appendix B (ranging from -1 to +2). To generate the simulated DC data, we randomly assigned a price level out of all price levels and simulated a consumer's response to the DC question depending on her generated actual WTP.

We repeated the above steps for 10 different sets of price levels. We started with 21 price levels ranging from CHF 0 to CHF 100 incremented in steps of CHF 5 (i.e., 0, 5, 10, 15, 20, 25, 30, 35, 40, 45, 50, 55, 60, 65, 70, 75, 80, 85, 90, 95, 100). In a second step, we removed the CHF 0 and CHF 100 levels. In a third step, we further removed the CHF 0 and CHF 100, and the CHF 5 and CHF 95 price levels, and so on until we narrowed down the DC price range to only three remaining levels, CHF 45, CHF 50, and CHF 55. For each resulting set of price levels, we randomly drew 1,000 WTP samples for each simulated group OE, DC, and BDM. Importantly, the price levels were randomly allocated to DC WTP values in each set. This allowed keeping the number of respondents constant despite having fewer price levels (i.e., when there are fewer price levels they are simply randomly distributed to more individuals). We then applied the bias correction approaches BASIC, EPSILON and FULL, and calculated mean WTP, the optimal price, optimal quantity and profits including a 95%-confidence interval based on the 1,000 samples of the de-biased data. All results are depicted in Figures H.1, H.2.

***Results of Simulation.*** We distinguished two different cases for the range of the DC price levels: First, the DC price range could be wider than the price range of the actual WTP (i.e., to the left of the vertical line in Figures H.1, H.2), or narrower than the price range of the actual WTP (i.e., to the right of the vertical line in Figures H.1, H.2). If the DC range chosen is wider than the actual price range, both non-parametric and parametric versions of all three de-biasing approaches are able to recover the actual WTP values. If the DC range is narrower than the actual price range, we see that the de-biasing approaches will still be able to generate the actual values, as long as the range is not too narrow (i.e., the range is reduced down to seven price levels, which means that the DC range is 35% narrower than the actual price range). We see that if the DC range is not too narrow, the confidence intervals still overlap (> 7 price levels). If there are fewer price levels, the results start to change. For instance, the mean values significantly reduce. This can be explained by the non-parametric approach we used to calculate mean WTP for DC (Figure H.1). In this case, narrowing the DC range will result in missing observations for large values and a missing upper bound for WTP, which will be ignored and not count towards the non-parametric mean. The lower values still count as the lower bound for WTP is always zero. In turn, this leads to a reduction of the mean at narrower DC ranges.

---

[7] Note: We did not use our empirical data from Study 1 and Study 2 for the simulation study, since we would lose all observations at the price levels we drop from the analysis. That is, we cannot reassign study participants from discarded DC price levels to other DC price levels still included in the simulation study, since their actual WTP is unkown to us.

Alternatively, the DC mean can also be calculated using a parametric approach as follows: First, a logistic regression is estimated based on the price levels and yes/no responses. The estimated coefficients for the intercept and slope then define a parametric logistic function. This function covers the full range of possible prices, allowing an unbiased calculation of the DC mean (using $\frac{1}{-\beta} \times log(1 + e^{\alpha})$); see also Hanemann 1989). This is confirmed in figures H.2 where the parametric de-biasing approaches generated accurate mean WTPs, optimal price, optimal quantity, and optimal profit values over all ranges of DC price levels.

# FIGURE H.1
# SENSITIVITY OF NON-PARAMETRIC DE-BIASING APPROACHES TO A REDUCTION OF THE NUMBER OF DC PRICE LEVELS

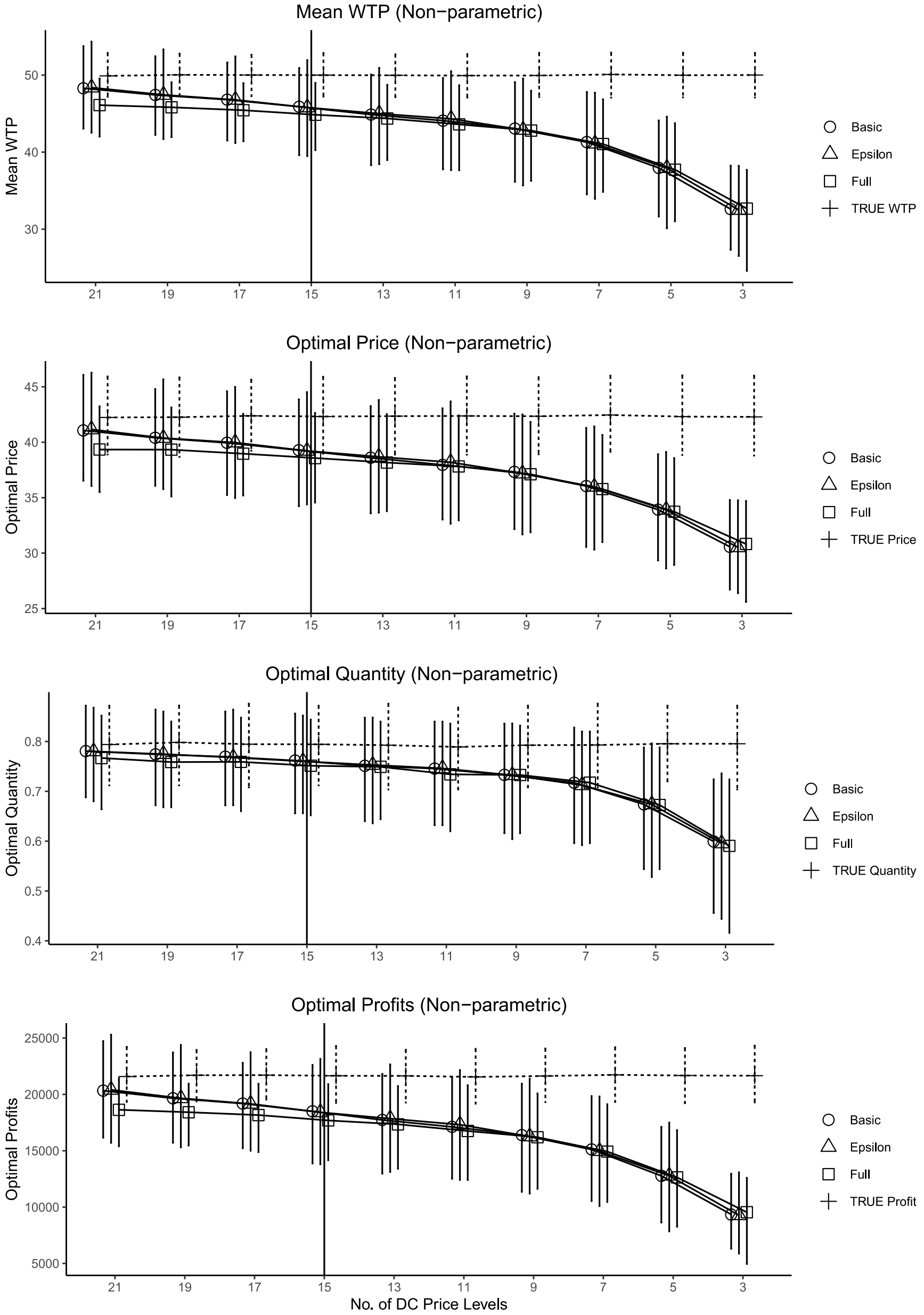

Note: We use a dodged presentation to facilitate interpretation of the graphs. Pluses indicate the true value of Mean WTP, Optimal Price, Quantity, and Profit. The circles, triangles and squares indicate the values recaptured by the respective non-parametric de-biasing approach, depending on the number of DC price levels used when collecting DC data (x-axis). The vertical line indicates the actual price range of the simulated WTP data. On the left of the vertical line, the additional price levels are outside the range of actual WTP values. On the right of the vertical line, the price levels are within the range of actual WTP values.

# FIGURE H.2
# SENSITIVITY OF PARAMETRIC DE-BIASING APPROACHES TO A REDUCTION OF THE NUMBER OF DC PRICE LEVELS

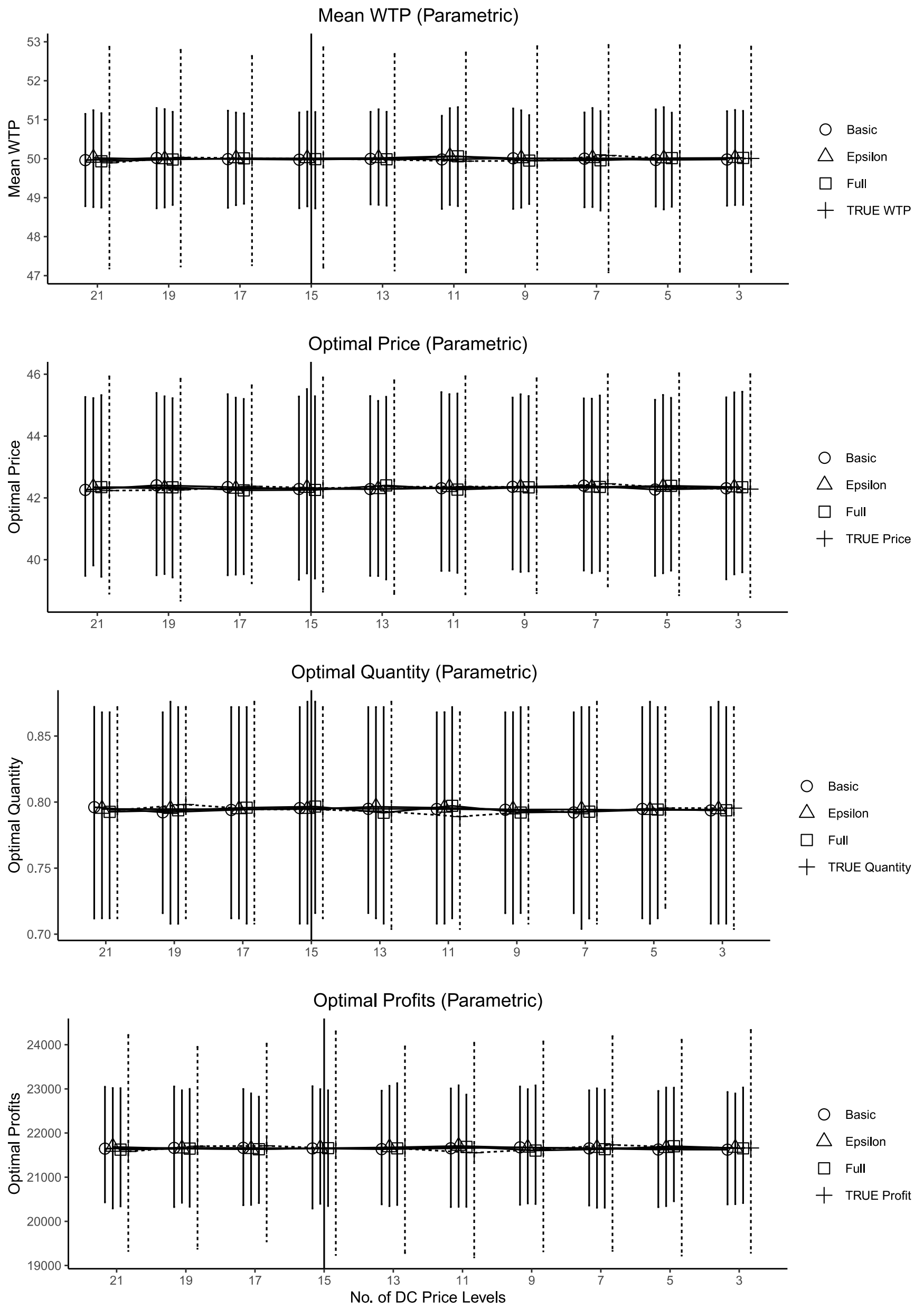

Note: We use a dodged presentation to facilitate interpretation of the graphs. Pluses indicate the true value of Mean WTP, Optimal Price, Quantity, and Profit. The circles, triangles and squares indicate the values recaptured by the respective parametric de-biasing approach, depending on the number of DC price levels used when collecting DC data (x-axis). The vertical line indicates the actual price range of the simulated WTP data. On the left of the vertical line, the additional price levels are outside the range of actual WTP values. On the right of the vertical line, the price levels are within the range of actual WTP values.

## *WEB APPENDIX I*

## *MANAGERIAL APPLICATION OF DE-BIASING APPROACH*

1. Managers interested in applying the de-biasing approach should first evaluates its applicability and validity to their particular business context and product or service. The discussion of alternative de-biasing approaches in the paper and Tables 1 and 2 provide a basis for such an evaluation. The de-biasing approach makes most sense when it is impossible or very costly to perform a price experiment with actual customers or use and incentive-aligned approach such as BDM.
2. OPTIONAL: If the BDM approach can be applied among a small sample of consumers (Note: In line with Simmons (2014), we recommend a minimum of 50), it can be used to capture the covariance used for the FULL de-biasing approach and to gauge the price range. The price range can be used to inform the price levels in the DC approach. Note that this step is not required for BASIC and EPSILON de-biasing.
3. Carefully define the range of DC prices and the number of levels in between based on the information gathered in step 2. If you did not perform step 2, use the cheapest and most expensive price levels in your particular market as low and high price levels for DC. When in doubt about the DC price levels, rather use a wider than narrower range. To reduce the sensitivity of the results to misspecifications of the price ranges, make sure to use a parametric approach in calculating the DC mean.
4. Collect OE and DC data using a between subjects design. Randomly allocate half of your participants to the DC question and half to the OE. Make sure that you collect a representative random sample of your population of interest (typically your target market).
5. Apply one of the de-biasing approaches BASIC, EPSILON, or FULL (in case you determined the covariance) as described in the paper.
6. Use de-biased demand functions for your pricing decision-making.